\documentclass[journal=jctcce,manuscript=article]{achemso}

\usepackage{amsmath}
\usepackage{bm}
\usepackage[version=3]{mhchem}
\usepackage{soul}
\usepackage{color}

\newcommand{\bra}[1]{\ensuremath{\langle \: #1 \: |}}
\newcommand{\ket}[1]{\ensuremath{| \: #1 \: \rangle}}

\newcommand{\OvI}[2]{\ensuremath{\langle \: #1 \mid #2 \: \rangle}}
\newcommand{\ExV}[3]{\ensuremath{\langle \: #1 \mid #2 \mid #3 \: \rangle}}
\newcommand{\OfPrn}[1]{\ensuremath{\left( #1 \right)}}
\newcommand{\OfPrnQ}[1]{\ensuremath{\left[ #1 \right]}}

\definecolor{lightblue}{rgb}{0.659,0.8706,1}
\definecolor{blu}{rgb}{0.2039,0.388,1}
\sethlcolor{lightblue}
\setstcolor{red}

\title{Large-scale quantum-dynamics with matrix product states}
\author{Alberto Baiardi}
\email{alberto.baiardi@phys.chem.ethz.ch}
\author{Markus Reiher}
\email{markus.reiher@phys.chem.ethz.ch}
\affiliation{ETH Z\"{u}rich, Laboratorium f\"{u}r Physikalische Chemie, Vladimir-Prelog-Weg 2, 8093 Z\"{u}rich, Switzerland.}
\date{\today}

\abbreviations{}
\keywords{Quantum dynamics, density matrix renormalization group, matrix product state, matrix product operator}

\begin{document}


\begin{abstract}

Dynamical electronic- and vibrational-structure theories have received a growing interest in the last years due to their ability to simulate spectra recorded with ultrafast experimental techniques. The exact time evolution of a molecular system can, in principle, be obtained from the time-dependent version of full configuration interaction. Such an approach is, however, limited to few-atom systems due to the exponential increase of its cost with the system dimension. In the present work, we overcome this unfavorable scaling by employing the time-dependent density matrix renormalization group (TD-DMRG) which parametrizes the time-dependent wavefunction as a matrix product state. The time-dependent Schr\"{o}dinger equation is then integrated with a sweep-based algorithm, as in standard time-independent DMRG. Unlike other TD-DMRG approaches, the one presented here leads to a set of coupled equations that can be integrated exactly. The resulting theory enables us to study real- and imaginary-time evolutions of Hamiltonians comprising more than 20 degrees of freedom that are challenging for current state-of-the-art quantum dynamics algorithms. We apply our algorithm to the simulation of quantum dynamics of models of increasing complexity, ranging from simple excitonic Hamiltonians to more complex \textit{ab-initio} vibronic ones.

\end{abstract}

\clearpage

\section{Introduction}
\label{sec:intro}

The vast majority of the algorithms in quantum chemistry aims at solving the time-independent Schr\"{o}dinger equation. 
A wide class of phenomena are, however, difficult to target within a time-independent framework. This is the case, for example, for the simulation of spectra obtained with time-resolved spectroscopic techniques that probe a molecular system under non-equilibrium conditions with pulsed light.\cite{Chergui2004_Review,Cho2008_Review,Mukamel2009_Review,Chergui2015_Account,Nisoli2017} Such spectra are readily obtained from a direct solution of the time-dependent Schr\"{o}dinger equation with quantum dynamics approaches.\cite{Nenov2014_TrackingPolipeptide2DUV,Santoro2017_VIPER-TDTI,Mukamel2017_Review,Cho2019_Coherent-2DIR_Review} Similarly, high-order response properties or high-lying excited states can, in principle, be obtained with time-independent approaches, but their calculation is easier within a time-dependent framework.\cite{Saalfrank2007_Properties-RealTime,Ding2013_Polarizability-RealTimeDFT,Li2018_RealTime-GUGA_CI}

Multiconfigurational time-dependent Hartree (MCTDH)\cite{Beck2000_MCTDH} is currently the reference method for quantum-dynamics simulations with vibrational\cite{Worth2008_MCTDH-Review,Meyer2011_MCTDH-Review} and electronic Hamiltonians.\cite{Saalfrank2005_MCTDH-Electrons,Cederbaum2012_MCTDH-Fermions} 
As any full-configuration interaction (full-CI) approach, MCTDH suffers from the curse of dimensionality because its computational cost increases exponentially with system size. This unfavorable scaling can be partially overcome with multi-scale formulations of MCTDH, such as the Gaussian MCTDH (G-MCTDH)\cite{Burghardt2008_VibronicMCTDH,Burghardt2013_ML-G-MCTDH} or the multi-layer MCTDH (ML-MCTDH).\cite{Manthe2008_MLMCTDH-Original,Wang2009_SQMCTDH,Manthe2011_MLMCTDH-Malonaldehyde,Vendrell2011_ML-MCTDH,Meyer2013_MLMCTDHVibronic,Wang2015_ML-MCTDH} In G-MCTDH, only a subset of the nuclear degrees of freedom are treated with conventional MCTDH, while the rest is described semiclassically with Gaussian wavepackets. Conversely, in ML-MCTDH, strongly correlated vibrational degrees of freedom are first contracted together with MCTDH. The resulting pre-contracted basis is truncated and then employed in a second MCTDH calculation. Systems with several hundreds of vibrational degrees of freedom can be targeted by this hierarchical treatment.\cite{Meyer2013_MLMCTDHVibronic,Burghardt2013_ChargeSeparation-MLMCTDH,Xie2015_ML-MCTDH-C60,Kuhn2016_ML-MCTDH-FMO} The efficiency of such multi-layer schemes depends on the choice and definition of the layers, \textit{i.e.,} on the choice of the modes treated with conventional MCTDH for G-MCTDH, and on the precontraction scheme for ML-MCTDH. Well-established schemes to perform this selection are, however, not yet available.

The density matrix renormalization group (DMRG) algorithm has become one of the reference methods for variational calculations on large systems in recent years. Unlike full-CI, the scaling of DMRG is polynomial in the system size and therefore does not suffer from the curse of dimensionality.\cite{White1992_DMRGBasis,White1993_DMRGBasis,Schollwoeck2005,Schollwoeck2011_Review-DMRG} In its original formulation, DMRG is designed for targeting ground states of Hamiltonians of linear systems with nearest-neighbor interactions. It has been then successfully applied to electronic\cite{Chan2011,Keller2014,Yanai2015,Olivares2015_DMRGInPractice,Szalay2015_Review} and vibrational\cite{Oseledts2016_VDMRG,Baiardi2017_VDMRG,Baiardi2019_HighEnergy-vDMRG,Roy2018_RotationalDMRG} Hamiltonians. DMRG replaces the exact diagonalization of the Hamiltonian of an $N$-body system with a series of smaller eigenvalue problems. Within each problem, only one site (\textit{i.e.}, an orbital for the electronic Hamiltonian) is treated explicitly and the effects of the remaining sites are represented through an effective basis, known as renormalized basis that is iteratively constructed to provide the best representation of the overall wavefunction in a least-squares sense. The dimension of this basis, also known as bond dimension, determines the accuracy of DMRG.

The generalization of DMRG to time-dependent processes faces two problems. The first one concerns the update of the renormalized basis during the propagation. The renormalized basis is known only at the initial time $t$ and must be optimized iteratively to optimize the representation of the wave function over the whole time interval. In the first TD-DMRG formulation, this basis was constructed at the beginning of the simulation and kept constant during the propagation.\cite{Cazalilla2002_TD-DMRG_FixedBasis} Not surprisingly, such a choice leads to a continuous decrease of accuracy with increasing time, which makes it useful for short-time propagations only.\cite{Luo2002_Comment-TDDMRG} More recent TD-DMRG formulations, known as adaptive TD-DMRG,\cite{Feiguin2005_Adaptive-TDDMRG,Feguin2006_Adaptive-TDDMRG,GarciaRipoll2006_Adaptive-TDDMRG,Ronca2017_TDDMRG-Targeting} build the renormalized basis to provide a balanced  least-squares representation of the wavefunction for the whole time-step of the propagation.

For Hamiltonians containing only nearest-neighbor interactions, an efficient solution is to express the time evolution operator through a Suzuki-Trotter factorization\cite{Suzuki1976_TrotterApproximation} as a product of terms involving only one and two sites at a time. The time evolution operator can be applied in a sweep-based fashion as in time-independent DMRG. The resulting theory is known as time-evolving block decimation (TEBD).\cite{Vidal2004_TEBD,White2004_AdaptiveTD-DMRG} Despite its successful applications for model Hamiltonians, TEBD has strong limitations when applied to more complex Hamiltonians, such as the ones of interest in quantum chemistry that contain long-range Coulomb interactions (highlighted by 4-indexed integrals in its second quantization form).

Shortly after its introduction, it was realized that DMRG optimizes wavefunctions encoded as matrix product states (MPSs),\cite{Oestlund1995_MPS} a parametrization known as tensor train in the mathematical community.\cite{Oseledets2011_TTGeneral} The MPS form can be generalized to operators to be written in matrix product operator (MPO) format.\cite{McCulloch2007_FromMPStoDMRG,Frowis2010_MPOGeneric,Chan2016_MPO-MPS,Hubig2017_GenericMPO} The availability of an analytic parametrization for wavefunctions and operators enables one to simulate the time evolution with the so-called Dirac-Frenkel variational principle.\cite{Moccia1973_TDVP,VanLeuven1988_EquivalenceTDPrinciple} This leads to an approximate form of the time-dependent Schr\"{o}dinger equation that is projected onto the manifold formed by all MPSs with a given bond dimension.\cite{Haegeman2011_TDDMRG-MPSMPO} In practice, this is achieved by scaling the Hamiltonian, expressed as an MPO, by the so-called tangent-space projector.\cite{Holz2012_ALSTheory} For this reason, the resulting theory is known as tangent-space formulation of TD-DMRG.\cite{Holz2012_ALSTheory,Lubich2014_TimeIntegrationTT,Haegeman2016_MPO-TDDMRG} We note that other algorithms to propagate wavefunctions encoded as MPSs have been proposed. Greene and Batista introduced the tensor-train spit-operator Fourier Transform (TT-SOFT)\cite{Batista2017_TT-SOFT} theory, in which the MPS and the propagator are constructed from their representation in the position/momentum basis. More recently, Frahm and Pfannkuche extended the Runge-Kutta and Lanczos propagation schemes to wave functions encoded as MPS.\cite{Frahm2019_TD-DMRG_Ultrafast} Both approaches share, however, a common limitation because the compact structure of an MPS is not fully exploited and, therefore, the bond dimension increases during the propagation. To avoid an uncontrolled increase of its size, the MPS needs to be truncated after each time-step. This does not hold true for the tangent-space-based scheme that implicitly keep the dimension of the MPS fixed.

The efficiency of TD-DMRG depends on the possibility of representing the wavefunction as a compact MPS for the whole propagation. The area law\cite{Hastings2007_AreaLaw} ensures that such representation is possible for ground states of short-range, gapped Hamiltonians. A time-dependent analog of this theorem is, however, not known, and it has been demonstrated for some spin Hamiltonians\cite{Montangero2006_EntanglementIncrease,Pollman2012_UnboundedEntanglement-TD} that the entanglement entropy increases during a time-dependent propagation even with a dynamical update of the renormalized basis. The dimension of the MPS must then also increase during the propagation to represent the wavefunction with constant accuracy. For this reason, particular care is needed when assessing the convergence of TD-DMRG propagations with respect to the renormalized basis set size.

The tangent-space-based TD-DMRG algorithm has been mostly applied to spin Hamiltonians\cite{Verstraete2017_TD-Schwinger,Goto2019_LongTime-TDDMRG} and chemical applications have been limited to excitonic models.\cite{Kurashige2018_MPS-MCTDH,Tempelaar2018_Holstein-TDDMRG,Yao2018_AdaptiveTDDMRG} In the present work, we investigate this algorithm on realistic vibrational and vibronic Hamiltonians containing both short- and long-range interactions. Moreover, we improve the flexiblity of the algorithm by allowing for a dynamical selection of the dimension of the MPS, within either a single-site\cite{McCulloch2015_Mixing} and a two-site\cite{Haegeman2016_MPO-TDDMRG} scheme. First, we study the performance of our implementation on an excitonic Hamiltonian\cite{Ivanov2015_ExcitonicReview} and compare to alternative TD-DMRG approaches.\cite{Ren2018_TDDMRG-Temperature,Kurashige2018_MPS-MCTDH,Tempelaar2018_Holstein-TDDMRG} Then, we simulate the full-dimensional excited-state nuclear dynamics of pyrazine by treating all 24 modes at the quantum level. Due to its size, this system represents a challenge for full quantum dynamics approaches and usually requires multi-layer treatments. Finally, the flexibility of the approach will be demonstrated by extending it to the simulation of the imaginary-time propagation of MPSs. This results in an efficient algorithm to optimize ground and excited states encoded as MPSs.

\section{Theoretical framework}
\label{sec:theory}

\subsection{MPS/MPO formulation of DMRG}

Before presenting the TD-DMRG theory, we describe briefly the time-independent formulation of DMRG to introduce the relevant notation. A more detailed discussion can be found in several recent reviews.\cite{Schollwoeck2005,Schollwoeck2011_Review-DMRG,Chan2011,Szalay2015_Review}

DMRG is an iterative optimization algorithm to approximate the ground state wavefunction $\ket{\Psi_0}$ of an $L$-sites quantum system.
In its MPS/MPO formulation, the wavefunction is expressed as an MPS as\cite{Oestlund1995_MPS,McCulloch2007_FromMPStoDMRG}

\begin{equation}
| \Psi \rangle = \sum_{\sigma_1,...,\sigma_L} \sum_{a_1,...,a_\text{L-1}} 
{M}_{1,a_1}^{\sigma_1} {M}_{a_1,a_2}^{\sigma_2}...
{M}_{a_\text{L-1},1}^{\sigma_L}
| \sigma_1,...,\sigma_L \rangle \, .
\label{eq:MPS}
\end{equation}
where $| \sigma_1,...,\sigma_L \rangle = | \boldsymbol{\sigma} \rangle$ are many-particle basis set elements obtained from the direct product of the basis of each site and expressed as occupation number vectors (ONV). The $L$ tensors $\bm{M}^{\sigma_i} = \{ M_{a_\text{i-1},a_\text{i}}^{\sigma_i} \}$ (usually known as site tensors) have one index ($\sigma_i$) associated to the local basis of the $i$-th site, and two additional indices with maximum dimension $m$ (note that $\mathbf{M}^{\sigma_1}$ and $\mathbf{M}^{\sigma_L}$ are row and column vectors, respectively, of maximum size $m$). A full-CI wavefunction can, in principle, be expressed exactly as an MPS, but its dimension would increase exponentially with system size. The key advantage of DMRG is the fact that ground states of Hamiltonians containing short-range interactions can be encoded by an MPS whose dimension is largely independent of the system size.\cite{Hastings2007_AreaLaw} For more general Hamiltonians, with both short- and long-range interactions, MPSs with larger $m$ are usually required to obtain a converged representation of the wavefunction.

Following the site decomposition of an MPS, the matrix product operator (MPO) representation of an operator $\mathcal{W}$ is obtained as\cite{McCulloch2007_FromMPStoDMRG,Schollwoeck2011_Review-DMRG,Keller2015_MPSMPODMRG,Chan2016_MPO-MPS}

\begin{equation}
\mathcal{W} = \sum_{\bm{\sigma\sigma'}} \sum_{b_1,\dots,b_{L-1}}^{b_1^\text{max},\dots,b_{L-1}^\text{max}}
W_{1 b_1}^{\sigma_1,\sigma_1'} \dots W_{b_{i-1} b_i}^{\sigma_i,\sigma_i'} \dots W_{b_{L-1} 1}^{\sigma_L,\sigma_L'}
| \bm{\sigma} \rangle \langle \bm{\sigma'} | \, ,
\label{eq:MPO}
\end{equation}
where $\bm{W}^{\sigma_l,\sigma_l'} = \{ W_{b_{l-1},b_l}^{\sigma_l,\sigma_l'} \}$ are 4-dimensional tensors with two indices ($\sigma_l$ and $\sigma_l'$) associated to the local basis of the $i$-th site.

The energy functional obtained combining Eqs.~(\ref{eq:MPS}) and (\ref{eq:MPO}) is a complex, non-linear function of the entries $\{ M_{a_\text{i-1},a_\text{i}}^{\sigma_i} \}$. If the energy is minimized with respect to the $l$-th site tensor by keeping all the other ones fixed, we obtain\cite{Schollwoeck2011_Review-DMRG}

\begin{equation}
\sum_{\sigma_i'} \sum_{a_{i-1}',a_i'} \sum_{b_{i-1},b_i} 
W_{b_{i-1},b_i}^{\sigma_i,\sigma_i'} L_{a_{i-1},a_{i-1}'}^{b_{i-1}} R_{a_i,a_i'}^{b_l} M_{a_{i-1}',a_i'}^{\sigma_i'} =
E M_{a_{i-1},a_i}^{\sigma_i} \, .
\label{eq:EnergyMinimization_TI}
\end{equation}
where $\bm{L}$ and $\bm{R}$ are three-dimensional tensors usually defined as left and right boundary, respectively. The left boundary is obtained by contracting the tensor network representing $\ExV{\Psi}{\mathcal{H}}{\Psi}$ for all sites ranging from 1 to $i$-1. Similarly, the right boundary is the partial contraction of the MPS with the MPO for the sites ranging from $i$+1 to $L$. Eq.~(\ref{eq:EnergyMinimization_TI}) can be recast as a standard eigenvalue equation $\bm{H} \bm{v} = E \bm{v}$, where $\bm{v}$ is a vector collecting all the entries of the $M_{a_{l-1},a_l}^{\sigma_l}$ tensor. The matrix $\bm{H}$ is defined as

\begin{equation}
H_{ (a_{i-1} \sigma_i a_i) , (a_{i-1}' \sigma_i' a_i') }
= \sum_{b_{i-1}} \sum_{b_i}  L_{a_{i-1},a_{i-1}'}^{b_{i-1}} W_{b_{i-1},b_i}^{\sigma_i,\sigma_i'} 
R_{a_i,a_i'}^{b_i} \, .
\label{eq:HamiltonianLocalRepresentation}
\end{equation}

In DMRG iterations, Eq.~(\ref{eq:EnergyMinimization_TI}) is solved sequentially for each site, starting from the first site up to the end of the chain (forward sweep) and then repeating this procedure with the reverse direction (backward sweep). By consistently following the lowest energy eigenpair, an approximation of the ground state is iteratively constructed.

\subsection{MPS/MPO formulation of TD-DMRG}

To apply the DMRG optimization algorithm of MPS in quantum dynamics simulations, the time-dependent Schr\"{o}dinger equation,
\begin{equation}
\mathcal{H} | \Psi(t) \rangle = i \partial_t | \Psi(t) \rangle \, ,
\label{eq:TDSE}
\end{equation}
must be solved under the condition that the wavefunction $| \Psi(t) \rangle$ is an MPS at any time $t$. It follows from Eq.~(\ref{eq:TDSE}) that the variation of the wavefunction $| \Psi (t) \rangle $ over a time-step $\Delta t$ is proportional to $\mathcal{H} | \Psi \rangle \times \Delta t$. It is known that, if $| \Psi (t) \rangle $ is represented as MPS with bond dimension $m$, an exact representation of $\mathcal{H} | \Psi \rangle$ will require an MPS with bond dimension $b_i m$, where b$_i$ is the bond dimension of the MPO (see Eq.~(\ref{eq:MPO})) and $m$ is the bond dimension of the MPS (see Eq.~(\ref{eq:MPS})). A naive integration of Eq.~(\ref{eq:TDSE}) would, therefore, lead to an unbounded increase of the MPS bond dimension.

For approximate wavefunction expressed by some parametrization scheme, Eq.~(\ref{eq:TDSE}) cannot be solved exactly in the most general case. The best approximation to the exact wavefunction $| \Psi(t) \rangle$ is obtained from the minimization of the Dirac-Frenkel functional F(t) defined as\cite{Moccia1973_TDVP,VanLeuven1988_EquivalenceTDPrinciple}

\begin{equation}
F(t) = \left\| \mathcal{H} \ket{\Psi} - i \partial_t \ket{\Psi} \right\|^2 \, ,
\label{eq:TD_Functional}
\end{equation}
where the minimization is performed within the set $S$ of all wavefunctions that can be expressed with the parametrization encoding $| \Psi(t) \rangle$. Eq.~(\ref{eq:TD_Functional}) is equivalent to

\begin{equation}
\langle \delta \Psi | \mathcal{H} - i \partial_t | \Psi \rangle = 0 \, ,
\label{eq:TDVP}
\end{equation}
where $| \delta \Psi \rangle$ is a generic infinitesimal variation of the wavefunction $| \Psi \rangle$ in $S$. The best approximation of the exact, time-dependent wavefunction $| \Psi(t) \rangle$ is therefore orthogonal, at each time step, to the error with respect to the exact solution of the time-dependent Schr\"{o}dinger equation $(\mathcal{H} - i \partial_t) | \Psi \rangle$.

Let now $\mathcal{M}_\text{MPS}(m)$ be the set of all MPSs with bond dimension $m$. It was proven\cite{Holz2012_ALSTheory} that $\mathcal{M}_\text{MPS}(m)$ is a smooth submanifold of the full Hilbert space. This means, in practice, that it is a subset of the full Hilbert space not displaying discontinuities and this facilitates the approximation of $\mathcal{M}_\text{MPS}(m)$ in the vicinity of this MPS as a linear subspace of the full configurational space. This linear approximation is known as the tangent space to $\mathcal{M}_\text{MPS}(m)$ calculated for the reference MPS $\ket{\Psi_\text{MPS}}$. Eq.~(\ref{eq:TDVP}) can be conveniently expressed in terms of the projector $\mathcal{P}_{\Psi_\text{MPS}}$ onto this tangent space,

\begin{equation}
i \partial_t | \Psi_\text{MPS} \rangle = \mathcal{P}_{\Psi_\text{MPS}} \mathcal{H} | \Psi_\text{MPS} \rangle \, .
\label{eq:TDVP_Projection}
\end{equation}

$\mathcal{P}_{\Psi_\text{MPS}}$ has a non-trivial effect on the right-hand side of Eq.~(\ref{eq:TDVP_Projection}) because, as already highlighted, the bond dimension of $\mathcal{H} \ket{\Psi_\text{MPS}}$ is larger than the one of $\ket{\Psi_\text{MPS}}$, $m$. The projection operation ensures that the dimension of the MPS is kept constant during the propagation.

A closed-form expression for $\mathcal{P}_{\Psi_\text{MPS}}$ was derived recently\cite{Holtz2012_ManifoldTT,Lubich2014_TimeIntegrationTT} and applied to the solution of Eq.~(\ref{eq:TDVP_Projection}).\cite{Haegeman2016_MPO-TDDMRG} We recall that an MPS is defined as canonically normalized with respect to site $l$ if it is represented as

\begin{equation}
| \Psi_\text{MPS} \rangle = \sum_{\sigma_1,...,\sigma_L} \sum_{a_1,...,a_\text{L-1}} 
{A}_{1,a_1}^{\sigma_1} \ldots {M}_{a_{l-1},a_l}^{\sigma_l} \ldots
{B}_{a_\text{L-1},1}^{\sigma_L} | \sigma_1, \ldots ,\sigma_L \rangle \, .
\label{eq:MPS_Canonical}
\end{equation}
where $\bm{A}$ and $\bm{B}$ fulfill the following orthogonality relation,

\begin{align}
\sum_{\sigma_l,a_{l-1}} {A}_{a_{l-1},a_l}^{\sigma_l} {A}_{a_{l-1},a_l'}^{\sigma_l} &= \delta_{a_l,a_l'} \\
\sum_{\sigma_l,a_l} {B}_{a_{l-1},a_l}^{\sigma_l} {B}_{a_{l-1}',a_l}^{\sigma_l} &= \delta_{a_{l-1},a_{l-1}'}
\label{eq:LeftRight_normalized}
\end{align}

An element $| \Psi_T \rangle$ of the tangent space to an MPS $| \Psi_\text{MPS} \rangle$ of a given rank $m$ can be decomposed as,\cite{Lubich2014_TimeIntegrationTT}

\begin{equation}
| \Psi_T \rangle = \sum_{i=1}^L | \Psi_T^{(i)} \rangle \, ,
\label{eq:MPS_Tangent_parametrization}
\end{equation}
where,

\begin{equation}
\begin{aligned}
| \Psi_T^{(i)} \rangle = \sum_{\sigma_1,...,\sigma_L} \sum_{a_1,...,a_\text{L-1}} 
{A}_{1,a_1}^{\sigma_1} \ldots & {N}_{a_{i-1},a_i}^{\sigma_i} \ldots
{B}_{a_\text{L-1},1}^{\sigma_L} | \sigma_1, \ldots ,\sigma_L \rangle \, .
\end{aligned}
\label{eq:MPS_Tangent_parametrization_element}
\end{equation}
$\bm{A}$ and $\bm{B}$ are given in Eq.~(\ref{eq:MPS_Canonical}) and ${N}_{a_{i-1},a_i}^{\sigma_i}$ is a normalized tensor orthogonal to ${M}_{a_{i-1},a_i}^{\sigma_i}$ introduced in Eq.~(\ref{eq:MPS_Canonical}). The orthogonality constraint ensures that the parametrization of Eq.~(\ref{eq:MPS_Tangent_parametrization_element}) is non-redundant. The projector $\mathcal{P}_{\Psi_\text{MPS}}$ may be written as,

\begin{equation}
\begin{aligned}
\mathcal{P}_{\Psi_\text{MPS}} = \sum_{i=1}^{N} &
\sum_{a_{i-1}^{(l)},\sigma_i,a_i^{(r)}}
| a_{i-1}^{(l)} \sigma_i a_i^{(r)} \rangle \langle a_{i-1}^{(l)} \sigma_i a_i^{(r)} | 
- \sum_{i=1}^{N-1} \sum_{a_i^{(l)},a_i^{(r)}} 
| a_i^{(l)} a_i^{(r)} \rangle \langle a_i^{(l)} a_i^{(r)} | \\
\equiv \sum_{i=1}^N &  \mathcal{P}_i^{(1)} - \sum_{i=1}^{N-1} \mathcal{P}_i^{(2)}  \, ,
\end{aligned}
\label{eq:ProjectorExpression}
\end{equation}

with the left ($| a_{i-1}^{(l)} \rangle$) and right ($| a_{i}^{(r)} \rangle$) renormalized bases on site $i$ defined as

\begin{equation}
\begin{aligned}
| a_{i-1}^{(l)} \rangle &= \sum_{\sigma_1,\ldots,\sigma_{i-1}} \sum_{a_1^{(l)},\ldots,a_{i-2}^{(l)}}
A_{1,a_1^{(l)}}^{\sigma_1} \ldots  
A_{a_{i-2}^{(l)},a_{i-1}^{(l)}}^{\sigma_{i-1}}
| \sigma_1 , \ldots , \sigma_{i-1} \rangle \, , \\
| a_i^{(r)} \rangle     &= \sum_{\sigma_{i+1},\ldots,\sigma_L} \sum_{a_{i+1}^{(r)},\ldots,a_L^{(r)}}
B_{a_i^{(r)},a_{i+1}^{(r)}}^{\sigma_{i+1}} \ldots  
B_{a_L^{(r)},1}^{\sigma_L}
| \sigma_{i+1} , \ldots , \sigma_L \rangle \, . \\
\end{aligned} 
\label{eq:RenormalizedStates_Def}
\end{equation}
Eq.~(\ref{eq:ProjectorExpression}) can be rationalized as follows. The first term is the sum of projectors onto the $| a_{i-1}^{(l)} \rangle \otimes | \sigma_i \rangle \otimes | a_i^{(r)} \rangle$ basis for each site $i$. The expression of an MPS normalized with respect to site $i$ in this product basis is

\begin{equation}
\ket{\Psi_\text{MPS}} = \sum_{a_{i-1}^{(l)} \sigma_i a_i^{(r)}} M_{a_{i-1}^{(l)},a_i^{(r)}}^{\sigma_i}
| a_{i-1}^{(l)} \sigma_i a_i^{(r)} \rangle \, .
\label{eq:MPS_RenormalizedBasis}
\end{equation}

The MPS can be expanded as in Eq.~(\ref{eq:MPS_RenormalizedBasis}) for any site $i$ and this might suggest that the union of all product bases $| a_{i-1}^{(l)} \sigma_i a_i^{(r)} \rangle$ constitutes a complete basis for a rank-$m$ MPS. However, this basis set would be overcomplete. To show that, consider a left sweep from the $i$-th to the ($i$+1)-th site. The left renormalized basis for the ($i$+1)-th site ($\ket{a_{i}^{(l)}}$) can be expressed as a function of $\ket{a_{i-1}^{(l)}}$

\begin{equation}
| a_i^{(l)} \rangle = \sum_{a_{i-1}^{(l)},\sigma_i} A_{a_{i-1}^{(l)},a_i^{(l)}}^{\sigma_i} 
| a_{i-1}^{(l)} \sigma_i \rangle \, .
\label{eq:Relation_between_RB}
\end{equation}

$A_{a_{i-1}^{(l)},a_i^{(l)}}^{\sigma_i}$ is obtained by singular value decomposition (SVD) or QR factorization of the optimized tensor $M_{a_{i-1}^{(l)},a_i^{(l)}}^{\sigma_i}$. Eq.~(\ref{eq:Relation_between_RB}) highlights that $| a_i^{(l)} \rangle$ and $| a_{i-1}^{(l)} \sigma_i \rangle$ are linearly dependent basis sets. Therefore, $| a_i^{(l)} \sigma_i a_i^{(r)} \rangle$ are contained in both the spaces spanned by $| a_{i-1}^{(l)} \sigma_i a_{i+1}^{(r)} \rangle$ and by $| a_i^{(l)} \sigma_i a_{i+1}^{(r)} \rangle$. Including only the first sum in Eq.~(\ref{eq:ProjectorExpression}) would, therefore, lead to a double counting of the $| a_{i-1}^{(l)} \sigma_i a_i^{(r)} \rangle$ basis functions. This effect is corrected by the second sum appearing in Eq.~(\ref{eq:ProjectorExpression}).

The formal solution of Eq.~(\ref{eq:TDVP_Projection}) is $e^{-i \mathcal{P}_{\Psi_\text{MPS}} \mathcal{H} t} | \Psi_\text{MPS}(0) \rangle$, but a closed form of the exponential operator $e^{-i \mathcal{P}_{\Psi_\text{MPS}} \mathcal{H} t}$ cannot be derived. As suggested in Ref.~\citenum{Haegeman2016_MPO-TDDMRG}, the exponential operator can be approximated by a first-order Lie-Trotter splitting as

\begin{equation}
e^{-i \mathcal{P}_{\Psi_\text{MPS}} \mathcal{H} t} \approx 
\prod_{i=1}^{N}   e^{-i \mathcal{P}_i^{(1)} \mathcal{H} t}
\prod_{i=1}^{N-1} e^{ i \mathcal{P}_i^{(2)} \mathcal{H} t} \, .
\label{eq:LieTrotter}
\end{equation}
where higher-order terms involving the commutator between $\mathcal{P}_i^{(1)}$ and $\mathcal{P}_i^{(2)}$ have been neglected. Although a Lie-Trotter decomposition is also employed in TEBD,\cite{Vidal2004_TEBD} the factorization introduced in Eq.~(\ref{eq:LieTrotter}) is much more general. On the one hand, TEBD factorizes the time-evolution operator as a product of exponentials, one for each term appearing in the Hamiltonian. Such factorization is, however, efficient only for nearest-neighbor Hamiltonians. On the other hand, the Lie-Trotter splitting in Eq.~(\ref{eq:LieTrotter}) is based on the expression of the projector $\mathcal{P}_{\Psi_\text{MPS}}$ as a sum of site terms, but the factorization of the Hamiltonian operator is avoided.

The order of application of the local operators appearing in Eq.~(\ref{eq:LieTrotter}) follows naturally from the structure of the MPS. The $| a_i^{(l)} \rangle$ basis is easily constructed from its equivalent of the previous site $| a_{i-1}^{(l)} \rangle$, and it is, therefore, convenient to apply the operators $\mathcal{P}_i^{(1)}$ and $\mathcal{P}_i^{(2)}$ one after the other for increasing values of $i$. The most appealing feature of this MPS/MPO-based formulation of TD-DMRG is that the resulting coupled differential equation can be integrated analytically. The representation of the Hamiltonian in the $| a_{i-1}^{(l)} \sigma_i a_i^{(l)} \rangle$ basis is given in Eq.~(\ref{eq:HamiltonianLocalRepresentation}). The differential equation obtained for the exponential operators containing $\mathcal{P}_i^{(1)}$ can then be expressed as

\begin{equation}
\frac{\text{d}}{\text{d}t} M_{a_{i-1}a_i}^{\sigma_i} = - \sum_{a_{i-1}', a_i', \sigma_i'}
{\rm i} H_{ a_{i-1} \sigma_i a_i , a_{i-1}' \sigma_i' a_i'} M_{a_{i-1}'a_i'}^{\sigma_i'} \, ,
\label{eq:DiffEq_tensor}
\end{equation}
which is a first-order, constant-coefficient linear differential equation that can be integrated exactly. The integration of the differential equation involving the $\mathcal{P}_i^{(2)}$ projector follows a similar strategy, once the representation of the Hamiltonian in the $| a_i^{(l)} a_i^{(r)} \rangle$ basis is available. We recall that the $| a_i^{(l)} \rangle$ basis is obtained from the $| a_{i-1}^{(l)} \sigma_i \rangle$ by QR factorization,

\begin{equation}
| a_i^{(l)} \rangle = \sum_{a_{i-1}^{(l)},\sigma_i} A_{a_{i-1}^{(l)}\sigma_i,a_i^{(l)}} 
\ket{ a_{i-1}^{(l)} \sigma_i } \, ,
\label{eq:QR_Factorization}
\end{equation}
where the $\bm{A}$ tensor, which is left normalized according to Eq.~(\ref{eq:LeftRight_normalized}), is defined as

\begin{equation}
M_{a_{i-1}^{(l)},a_i^{(r)}}^{\sigma_i} =  \sum_{a_i^{(l)'}} A_{a_{i-1}^{(l)},a_i^{(l)'}}^{\sigma_i} Q_{a_i^{(l)'},a_i^{(r)}} \, .
\label{eq:QR_Factorization_2}
\end{equation}

The matrix representation of the Hamiltonian in the $| a_i^{(l)} a_i^{(r)} \rangle$ basis is then

\begin{equation}
\begin{aligned}
H_{\left( a_i^{(l)} a_i^{(r)} , a_i^{(l)'} a_i^{(r)'} \right)} &= \ExV{a_i^{(l)} a_i^{(r)}}
{\mathcal{H}}
{a_i^{(l),'} a_i^{(r),'}} \\
& = \sum_{\tilde{a}_{i-1},\tilde{a}_{i-1}'} \sum_{\sigma_i, \sigma_i'} 
A_{\tilde{a}_{i-1},a_i}^{\sigma_i,\dagger} A_{\tilde{a}_{i-1}',a_i'}^{\sigma_i'}
\ExV{\tilde{a}_{i-1}^{(l)} \sigma_i a_i^{(r)}}
{\mathcal{H}} 
{\tilde{a}_{i-1}^{(l),'} \sigma_i' a_i^{(r),'}} \, .
\end{aligned}
\label{eq:MatrixRepresentation_H}
\end{equation}

By combining Eqs.~(\ref{eq:HamiltonianLocalRepresentation}) and (\ref{eq:MatrixRepresentation_H}) and considering that the left boundary for site $i$ can be obtained recursively from the one of site $i$+1 through the following recursive relation,\cite{Schollwoeck2011_Review-DMRG,Keller2015_MPSMPODMRG}

\begin{equation}
L_{a_i^{(l)},a_i^{(l)'}}^{b_l} = \sum_{\sigma_i,\sigma_i'} \sum_{a_{i-1},a_{i-1}'} \sum_{b_{i-1}}
A_{\tilde{a}_{i-1}^{(l)},a_i^{(l)}}^{\sigma_i,\dagger} 
A_{\tilde{a}_{i-1}^{(l)'},a_i^{(l)'}}^{\sigma_i'}
W_{b_{i-1},b_i}^{\sigma_i,\sigma_i'} L_{a_{i-1}^{(l)},a_{i-1}^{(l)'}}^{b_{i-1}} \, ,
\label{eq:Boundaries_recursive}
\end{equation}
it is easy to show that Eq.~(\ref{eq:MatrixRepresentation_H}) can be expressed in a more compact form as

\begin{equation}
H_{\left( a_i^{(l)} a_i^{(r)} , a_i^{(l)'} a_i^{(r)'} \right)}
= \sum_{b_l} L_{a_i^{(l)}, a_i^{(l)'}}^{b_l} R_{a_i^{(r)}, a_i^{(r)'}}^{b_l} \, .
\label{eq:HamiltonianLocalRepresentation_2}
\end{equation}

The representation of an MPS in the $| a_i^{(l)} a_i^{(r)} \rangle$ basis is obtained by combining Eqs.~(\ref{eq:MPS_Canonical}) and (\ref{eq:Relation_between_RB}),

\begin{equation}
\begin{aligned}
| \Psi_\text{MPS} \rangle &= \sum_{\sigma_1,\ldots,\sigma_L} \sum_{a_1^{(l)},\ldots,a_i^{(l)}} \sum_{a_i^{(r)},\ldots,a_\text{L-1}^{(r)}}
A_{1,a_1^{(l)}}^{\sigma_1} \ldots A_{a_{i-1}^{(l)},a_i^{(l)}}^{\sigma_i}  \\
&  \qquad \qquad \times Q_{a_i^{(l)},a_i^{(r)}} B_{a_i^{(r)},a_{i+1}^{(r)}}^{\sigma_{i+1}} \ldots B_{a_{L-1}^{(r)},1}^{\sigma_L}
| \sigma_1 \ldots \sigma_L \rangle \\ 
& \equiv \sum_{a_i^{(l)},a_i^{(r)}} Q_{a_i^{(l)},a_i^{(r)}} | a_i^{(l)} a_i^{(r)} \rangle \, .
\end{aligned}
\label{eq:Representation_RenormalizedBasis}
\end{equation}

Hence, the representation is given by the $\bm{Q}$ matrix, which is often referred to as zero-site tensor.\cite{Haegeman2016_MPO-TDDMRG}

\subsection{Single- and two-site integrators}

The theoretical framework introduced above assumes that the bond dimension of the initial wavefunction $\ket{\Psi(t=0)}$ is kept constant throughout the whole propagation time. This is a major limitation in practical applications of TD-DMRG because the bond dimension $m$ of the wavefunction at the beginning of the propagation is determined by the initial conditions. In Section~\ref{sec:applications}, we will study the nuclear dynamics of a molecule after photoexcitation to an electronically excited state. Potential energy surfaces are modeled with the harmonic approximation and, therefore, the initial state is the vibrational ground state of the lowest electronic state. In the occupation number vector representation, this is a basis vector ($\ket{00\cdots0}$) that corresponds to an MPS with $m$=1. A straightforward application of the propagation algorithm described above would, therefore, constrain the bond dimension of the MPS to one. Here, we propose two alternative strategies, adapted from the standard time-independent variant of DMRG, to alleviate this limitation allowing for a dynamical adaptation of $m$.

The first route is based on the subspace expansion algorithm originally devised for tensor trains\cite{Dolgov2014_ALSTheory,Dolgov2015_SubspaceExpansion} and later applied to single-site time-independent DMRG.\cite{McCulloch2015_Mixing} We will discuss the algorithm for a forward sweep because the extension to a backward sweep is trivial. After its optimization, an MPS tensor $M_{a_{i-1},a_i}^{\sigma_i}$ is reshaped as a matrix $M_{(a_{i-1}\sigma_i,a_i)}$ and $m'$ columns are added to the resulting matrix by increasing its bond dimension to $m+m'$. The MPS tensor on the next site $M_{a_i,a_{i+1}}^{\sigma_{i+1}}$ is reshaped as $M_{(a_i,a_{i+1}\sigma_{i+1})}$ and $m'$ null rows are added to match the new dimensions of the preceding tensor. This modification does not change the MPS because the rows added on the $i$-th site tensor are contracted with null columns on the ($i$+1)-th site tensor. Nevertheless, this step enlarges the basis $a_i^{(l)}$, and therefore, the optimization of the MPS for the ($i$+1)-th site will be performed in a larger vector space. The efficiency of the subspace expansion step depends strongly on the $m'$ vectors that are added to $M_{a_{i-1},a_i}^{\sigma_i}$ at the end of a microiteration step. In principle,\cite{Lubich2016_LowRank-SmallSingularValues} these vectors can be chosen randomly without altering the efficiency of the DMRG optimization. A more reliable alternative is to construct the vectors as\cite{McCulloch2015_Mixing}

\begin{equation}
\Delta M_{a_{i-1}\sigma_i,b_i a_i'} = \sum_{b_{i-1}} \sum_{a_{i-1}'} \sum_{\sigma_i'}
L_{a_{i-1}a_{i-1}'}^{b_{i-1}} M_{a_{i-1}'a_i'}^{\sigma_i'} W_{b_{i-1}b_i}^{\sigma_i \sigma_i'} \, ,
\label{eq:SubspaceExpansion}
\end{equation}

where the resulting matrix $\Delta \bm{M}$ is to be added to the $\bm{M}^{\sigma_i}$ tensor. Eq.~(\ref{eq:SubspaceExpansion}) can be intuitively motivated as follows. We have already mentioned that applying an operator $\mathcal{W}$ encoded as an MPO to an MPS increases the bond dimension of the MPS. In fact,

\begin{equation}
\begin{aligned}
\mathcal{W} \ket{\Psi} &= 
\sum_{b_1 \ldots b_{i-1}} \sum_{a_1 \ldots a_{i-1}} \sum_{\sigma_1' \ldots \sigma_L'}
W_{1b_1}^{\sigma_1,\sigma_1'} \cdots W_{b_L 1}^{\sigma_L,\sigma_L'}
M_{1a_1}^{\sigma_1'} \cdots M_{a_{L-1}1}^{\sigma_L'} \ket{\sigma_1', \ldots , \sigma_L'} \\
&= \sum_{a_1' \ldots a_{L-1}'} \sum_{\sigma_1' \ldots \sigma_L'} 
N_{1 a_1'} \cdots N_{a_{L-1}' 1} \ket{\sigma_1', \ldots , \sigma_L'} \, ,
\end{aligned}
\label{eq:MPOOnMPS}
\end{equation}
where the index $a_i'$ combines the indices $b_i$ and $a_i$, and

\begin{equation}
N_{a_{i-1}'a_i'}^{\sigma_i} = \sum_{\sigma_i'} W_{b_{i-1}b_i}^{\sigma_i \sigma_i'} M_{a_{i-1} a_i}^{\sigma_i'} \, .
\label{eq:MPOonMPS_2}
\end{equation}

Therefore, for a given site $i$, the left- and right-renormalized bases representing $\mathcal{H} \ket{\Psi}$ are larger than the ones of $\ket{\Psi}$. The subspace expansion algorithm enlarges the left-renormalized basis for $\ket{\Psi}$ with the left-renormalized basis of $\mathcal{H} \ket{\Psi}$. It is easy to demonstrate that the matrix defined in Eq.~(\ref{eq:SubspaceExpansion}) expresses this new additional basis in terms of the $\ket{a_{i-1}} \otimes \ket{\sigma_i}$ product basis.

A direct application of the subspace expansion algorithm would increase the bond dimension by a factor of $b_i$ per microiteration, where $b_i$ is the bond dimension of the MPO introduced in Eq.~(\ref{eq:MPOOnMPS}). To avoid an uncontrolled increase of $m$, it was suggested in Ref.~\citenum{McCulloch2015_Mixing} to scale the matrix $\Delta \bm{M}^{\sigma_i}$ by a parameter $\mu<1$ and then truncate $\bm{M}^{\sigma_i}$ with SVD. Even if adaptive schemes to select $\mu$ have been proposed,\cite{McCulloch2015_Mixing} in the present work $\mu$ is kept constant for $n_\mu$ sweeps and then set to 0 because already such simple choice provides reliable and converged results.

A second, alternative algorithm to dynamically control the bond dimension $m$ is the two-site optimizer,\cite{Schollwoeck2011_Review-DMRG,Keller2015_MPSMPODMRG} in which the energy is minimized with respect to the simultaneous variation of two consecutive sites. In practice, a two-site tensor is defined as,

\begin{equation}
T_{a_{i-1}a_{i+1}}^{\sigma_i \sigma_{i+1}}
= \sum_{a_i} M_{a_{i-1}a_i}^{\sigma_i} M_{a_ia_{i+1}}^{\sigma_{i+1}} \, ,
\label{eq:TwoSitesTensor}
\end{equation}
and the optimization is carried out for $\bm{T}^{\sigma_i \sigma_{i+1}}$ instead of $M_{a_{i-1}a_i}^{\sigma_i}$. To return to the MPS factorization, $\bm{T}$ is first reshaped as a matrix as $T_{(a_{i-1}\sigma_i,a_{i+1}\sigma_{i+1})}$ and then factorized by SVD,

\begin{equation*}
T_{(a_{i-1}\sigma_i, a_{i+1}\sigma_{i+1})} = \sum_{a_i'=1}^{m'} U_{(a_{i-1}\sigma_{i-1},a_i')} S_{a_i', a_i'} V_{a_{i+1}\sigma_i,a_i'} \, ,
\label{eq:SVD_TwoSites}
\end{equation*}
where $\bm{U}$ and $\bm{V}$ are orthogonal matrices and $\bm{S}$ is a diagonal matrix containing the singular values. The upper bound $m'$ to the sum over the index $a_i'$ is equal to the rank of $T_{(a_{i-1}\sigma_{i-1},a_{i+1}\sigma_i)}$ and can, in general, be different from the rank of the matrix before the optimization. To avoid an uncontrolled increase of the updated bond dimension $m'$, the SVD can be truncated. For $S_\text{keep}$ being the set of diagonal elements that are kept after the truncation, it is possible to demonstrate\cite{Legeza2003_DynamicalBlockState} that the error $\eta_\text{err}$ introduced in the approximation is given by

\begin{equation}
\eta_\text{err} = 1 - \sum_{i \in S_\text{keep}} S_{ii}^2 \, .
\label{eq:TruncationError}
\end{equation}

It is, therefore, possible to fix a threshold on the error $\eta_\text{thresh}$ and keep the minimal number of vectors in the SVD truncation so that the error is smaller than $\eta_\text{thresh}$. This idea is at the heart of the dynamical block state selection scheme\cite{Legeza2003_DynamicalBlockState} and can be extended also to time-dependent propagations.

\subsection{MPO representation of vibrational and vibronic Hamiltonians}

The efficiency of DMRG relies on the ability of representing the Hamiltonian as a compact MPO. Different algorithms have been proposed to build MPO representations of general operators.\cite{Pirvu2010_MPORepresentation,Frowis2010_MPOGeneric,Dolfi2014_ALPSProject,Keller2015_MPSMPODMRG,Hubig2017_GenericMPO} Among them, we chose the one introduced in Ref.~\citenum{Keller2015_MPSMPODMRG} for electronic structure problems which we recently generalized to vibrational Hamiltonians.\cite{Baiardi2017_VDMRG,Baiardi2019_HighEnergy-vDMRG} This algorithm requires as input parameter a representation of the Hamiltonian in terms of local operators. The MPO is then represented as a graph, where each local operator is associated with a different node. A single term in the Hamiltonian is represented by a connection between the vertices. The major advantage of this construction scheme over other ones available in the literature is its generality. In fact, it supports strings of second-quantized creation and annihilation operators with arbitrary length and is not limited to Hamiltonians with short-range interactions.

Even if TD-DMRG is not limited to any specific form of the Hamiltonian, for the present work we chose to focus on vibrational and vibronic Hamiltonians. We will turn to electronic Hamiltonians in future works. Regarding the former class, we already discussed in Ref.~\citenum{Baiardi2017_VDMRG} how to generate an MPO representation of the Watson Hamiltonian,

\begin{equation}
\mathcal{H}_\text{vib} =
-\frac{1}{2} \sum_{i=1}^{L} \frac{\partial^2}{\partial q_i^2} +  \mathcal{V}(\bm{q}) \, ,
\label{eq:WatsonHamiltonian}
\end{equation}
where Cartesian normal modes $\bm{q}$ have been used as a reference coordinate set and the potential operator $\mathcal{V}(\bm{q})$ has been expanded as a Taylor series in powers of $\bm{q}$. A second-quantization representation if Eq.~(\ref{eq:WatsonHamiltonian}) is obtained by applying the canonical quantization rules to the position operator $\hat{q}_i$ and its conjugate momentum $\hat{p}_i$,

\begin{equation}
\begin{aligned}
\hat{q}_i &= \frac{1}{2}  \left( \hat{b}_i^+ + \hat{b}_i  \right) \, , \\
\hat{p}_i &= \frac{i}{2}  \left( \hat{b}_i^+ - \hat{b}_i  \right) \, , \\
\end{aligned}
\label{eq:CanonicalQuantization}
\end{equation}
where $\hat{b}_i^+$ and $\hat{b}_i$ are bosonic creation and annihilation operators, respectively. Within this canonical quantization representation, the basis set is represented in terms of occupation number vectors (ONVs) $\ket{n_1, n_2 \ldots n_L}$ where $n_i$ labels the occupation number of mode $i$.

Vibronic Hamiltonians describe the coupled dynamics of nuclei and electrons and, therefore, their encoding as MPOs is less straightforward. A wavefunction including nuclear and electronic degrees of freedom can be parametrized as

\begin{equation}
\ket{\Psi(t)} 
= \sum_{\alpha=1}^{N_\text{ex}} C_{\alpha}^e(t) \, \ket{\alpha} \otimes \ket{\Psi_\text{vib}^\alpha} \, .
\label{eq:VibronicWavefunctionParametrization}
\end{equation}

where we label the electronic wavefunction obtained within the Born-Oppenheimer approximation at a given reference geometry as $\ket{\alpha}$. $\ket{\Psi_\text{vib}^\alpha}$ is the vibrational wavefunction associated to the nuclear motion on the $\alpha$-th potential energy surface. For each value of $\alpha$, $\ket{\Psi_\text{vib}^\alpha}$ can be expanded as linear combination of harmonic oscillator eigenfunctions for each degree of freedom $i$. In the ONV representation, Eq.~(\ref{eq:VibronicWavefunctionParametrization}) can be rewritten as

\begin{equation}
\begin{aligned}
\ket{\Psi(t)} 
&= \sum_{\alpha=1}^{N_\text{ex}} \sum_{n_1=0}^{N_\text{max}} \ldots \sum_{n_L=0}^{N_\text{max}}
C_{\alpha}^e(t) \, C_{n_1,\ldots,n_L}^v(t) \, \ket{\alpha} \otimes \ket{n_1,\ldots,n_L} \\
&= \sum_{\alpha_1=0}^{1} \ldots \sum_{\alpha_{N_\text{ex}}=0}^{1} 
\sum_{n_1=0}^{N_\text{max}} \ldots \sum_{n_L=0}^{N_\text{max}}
C_{\alpha_1,\ldots,\alpha_{N_\text{ex}},n_1,\ldots,n_L}(t) \, 
\ket{\alpha_1, \ldots, \alpha_{N_\text{ex}}} \otimes \ket{n_1,\ldots,n_L} \, .
\end{aligned}
\label{eq:VibronicWavefunctionParametrization2}
\end{equation}

In the first line of Eq.~(\ref{eq:VibronicWavefunctionParametrization2}), the electronic degrees of freedom are labeled through an ONV with a single index, $\ket{\alpha}$, as in Eq.~(\ref{eq:VibronicWavefunctionParametrization}), where different electronic states are associated to different values for $\alpha$. In the second line, a separate list of indices $\ket{\alpha_1, \ldots, \alpha_{N_\text{ex}}}$ is employed for each electronic state in the ONV representation. The indices $\alpha_i$ can be either 0, if the molecule is not in electronic state $i$, or 1, if electrons are solely described by the $i$-th electronic wavefunction. Therefore, the following constraint must be fulfilled

\begin{equation}
\sum_{i=1}^{N_\text{ex}} \alpha_i = 1 \, .
\label{eq:Symm_Vibronic}
\end{equation}

Eq.~(\ref{eq:Symm_Vibronic}) reflects the conservation of the number of excited electronic states (usually known as excitons) and induces a block structure of the MPS. This structure can be exploited to reduce the computational cost of DMRG.\cite{Vidal2011_DMRG-U1Symm} 

The MPS form of the expansion given in the second line of Eq.~(\ref{eq:VibronicWavefunctionParametrization2}) is obtained as,

\begin{equation}
C_{\alpha_1,\ldots,\alpha_{N_\text{ex}},n_1,\ldots,n_L} 
= \sum_{a_1=1}^m \ldots \sum_{a_{L+N_\text{ex}-1}=1}^m 
M_{1,a_1}^{\alpha_1} M_{a_1,a_2}^{\alpha_2} \cdots M_{N_\text{ex}+L-1,1}^{n_L}
\label{eq:MPS_Vibronic}
\end{equation}

In principle, it is possible to derive an MPS representation of the first line of Eq.~(\ref{eq:VibronicWavefunctionParametrization2}). However, the resulting parametrization would be less flexible. As already discussed in the literature, the efficiency of DMRG depends on the sorting of the sites on the lattice and different strategies to derive an optimal sorting have been proposed\cite{Chan2002_DMRG,Legeza2003_OrderingOptimization,Moritz2005_OrbitalOrdering,Rissler2006_QuantumInformationOrbitals} of which the ordering according to orbital entropies has been demonstrated to be very efficient. In Eq.~(\ref{eq:MPS_Vibronic}), each electronic state is associated to a site and it is, therefore, possible to order the sites so that the electronic states are not close to each other. This is not possible in the first representation given in the first line of Eq.~(\ref{eq:VibronicWavefunctionParametrization2}), in which only one site is associated to the electronic degrees of freedom of the molecule.

This second, more flexible occupation number vector representation of the electronic degrees of freedom can be associated to second-quantizated creation and annihilation operators $\hat{a}_{\alpha}^+$ and $\hat{a}_{\alpha}$, respectively, such that

\begin{equation}
\begin{aligned}
\hat{a}_i^+ \ket{\ldots \alpha_i \ldots}  &= \delta_{\alpha_i,0} \ket{\ldots \alpha_i+1 \ldots} \\
\hat{a}_i   \ket{\ldots \alpha_i \ldots}  &= \delta_{\alpha_i,1} \ket{\ldots \alpha_i-1 \ldots} \, , \\
\end{aligned}
\label{eq:ElectronicSecondQuantization}
\end{equation}

The operators defined in Eq.~(\ref{eq:ElectronicSecondQuantization}) satisfy fermionic anticommutation rules but, due to the constraint of Eq.~(\ref{eq:Symm_Vibronic}), only one electronic site is occupied. Hence, no antisymmetry constraint must be imposed. For cases in which multiple excitations are present, the fermionic symmetry can be included as discussed, for example, in Ref.~\citenum{Keller2015_MPSMPODMRG}.

\section{Results}
\label{sec:applications}

We apply the TD-DMRG theory developed here to three examples. The first one is an aggregate of perylene bisimide-based dye molecule, whose excited-state dynamics can be modeled with an excitonic Hamiltonian. This molecular system has been already studied with both MCTDH\cite{Kuhn2012_ExcitonicPerylene-6Mode,Kuhn2013_NonAdiabatic-Excitonic,Ivanov2015_ExcitonicReview} and DMRG,\cite{Ren2018_TDDMRG-Temperature} and therefore, it is a suitable starting point for analyzing our TD-DMRG implementation against alternative formulations.

The second molecule studied in this work is pyrazine. The excited state dynamics for the S$_1$ $\leftarrow$ S$_2$ non-radiative decay simulated on the potential energy surface of Ref.~\citenum{Raab1999_PyrazineDiabatic} is the most common test case for new quantum dynamical approaches. Among the various TD-DMRG formulations proposed in the literature, only the TT-SOFT algorithm\cite{Batista2017_TT-SOFT} presented in the introduction has been applied, up to now, to pyrazine. A full-dimensional quantum dynamics study of pyrazine, performed by treating all 24 modes at the same level of accuracy, is challenging for most state-of-the-art approaches. We show here how TD-DMRG allows us to obtain converged spectra for this system by treating all vibrational degrees of freedom at the same footing without the need of introducing any hierarchical treatment of the modes, as is done in ML-MCTDH. The pyrazine example will also demonstrate the reliability of DMRG in targeting general vibronic Hamiltonians lacking a regular interaction pattern.

Finally, we extend TD-DMRG to imaginary time propagations to optimize ground and excited vibrational levels of ethylene. Imaginary-time propagation is the idea underlying a diffusion Monte Carlo approach and has been employed to optimize ground and excited states of electronic\cite{AspuruGuzik2015_ImaginaryTime} and vibrational\cite{Meyer2006_MCTDH-ImaginaryTime,Gatti2008_ImaginaryTime-Fluoroform,Manthe2012_Malonaldehyde-ImaginaryTime} Hamiltonians.

\subsection{Computational details}

We implemented the tangent-space-based TD-DMRG theory in our \textsc{QCMaquis-V} program designed to perform DMRG calculations on vibrational Hamiltonians\cite{Baiardi2017_VDMRG,Baiardi2019_HighEnergy-vDMRG} and derived from the \textsc{QCMaquis} program for electronic structure calculations.\cite{Keller2014,Keller2015_MPSMPODMRG}

If not otherwise specified, TD-DMRG propagations were performed with a time step of 1~fs and an overall propagation time of 4000 fs. Single-site simulations employ the subspace expansion variant presented in Ref.\citenum{McCulloch2015_Mixing}. The scaling factor $\mu$ for the MPS $\Delta \bm{M}$ (recall Eq.(\ref{eq:SubspaceExpansion} for the definition of $\Delta \bm{M}$) was set to 10$^{-8}$ for the first 100 sweeps (note that each sweep corresponds to one time step), and to 10$^{-15}$ up to the 200-th sweep, and then set to 0. Other combinations of values for $\mu$ and numbers of sweeps have been tested and provide results equivalent to the ones reported in the next section.

For excited-state nuclear dynamics, vibronic spectra were obtained from the Fourier transformation of the autocorrelation function $C(t) = \OvI{\Psi(t)}{\Psi(0)}$. Broadening effects were reproduced by scaling the autocorrelation function by an exponential damping function $f(t) = e^{-t/\tau}$, corresponding to a Lorentzian broadening function in the frequency domain. The parameter $\tau$ determines the decay rate of the autocorrelation function, and therefore, the broadening of the spectrum in the frequency domain. Low values of $\tau$ produce large broadenings, whereas large values of $\tau$ show sharp spectra.

The parameters of the excitonic Hamiltonian describing the energy transfer in perylene bisimide aggregates were taken from Ref.~\citenum{Ivanov2015_ExcitonicReview}. The parameters for the S$_1$/S$_2$ vibronic Hamiltonian of pyrazine were taken from Ref.~\citenum{Raab1999_PyrazineDiabatic}, employing the data calculated with multi-reference CI. All spectra were shifted by 35609~cm$^{-1}$ to match the experimental vertical excitation energy. As in our previous works,\cite{Baiardi2017_VDMRG,Baiardi2019_HighEnergy-vDMRG} the potential energy surface for ethylene was taken from Ref.~\citenum{Delahaye2014_EthylenePES} by expressing it as a quartic force-field in terms of the Cartesian-based normal modes employing the \texttt{PyPES} library.\cite{Crittenden2015_PyPES}

\subsection{Excitonic dynamics of perylene-bisimide aggregates}

We first apply our TD-DMRG theory to the simulation of the photochemical properties of a perylene bisimide-based dye, namely PBI-I. Perylene-bisimide derivatives can form molecular aggregates through $\pi$-stacking interactions. Upon absorption of UV light, the excitation is transferred efficiently between the constituents of the aggregate. As models for light-harvesting compounds, PBI-I aggregates have been extensively studied both computationally\cite{Meyer2009_MCTDH-Perylene,Kuhn2012_ExcitonicPerylene-6Mode,Kuhn2013_NonAdiabatic-Excitonic,Canola2018_Excitonic} and experimentally.\cite{Wurthner2004_PerileneBisimide,Wurthner2011_ExcitonicQuenching}

PBI-I features 400 vibrational modes and a full quantum treatment is prohibitive already for the isolated monomer in its electronic ground state. A simplified form of the vibronic Hamiltonian of molecular aggregates is obtained within the excitonic approximation. First, the electronic states of the aggregate are described within a diabatic representation, in which the diabatic states of the aggregate are assumed to coincide with the electronic states of the isolated monomers. Then, couplings between vibrations localized on different monomers are neglected. In the following, we label the ground state of the $i$-th monomer as $\alpha_0^{(i)}$ and its $n$-th excited state as $\alpha_n^{(i)}$. The vibronic Hamiltonian can be modeled as the so-called Holstein Hamiltonian

\begin{equation}
\mathcal{H}_\text{vib} = \ket{\bm{0}} \bra{\bm{0}} 
\OfPrnQ{ \sum_{i=1}^{N_\text{mol}} \Bigl( \mathcal{T} \OfPrn{\bm{q}_i} 
	+ \mathcal{V}_i^\text{GS} \OfPrn{\bm{q}_i} \Bigr) }
+ \sum_{\alpha=1}^{N_\text{exc}} \sum_{i=1}^{N_\text{mol}}  \ket{\bm{\alpha}_i} \bra{\bm{\alpha}_i}
\OfPrnQ{ \Bigl( \mathcal{T} \OfPrn{\bm{q}_i} + \mathcal{V}_i^{\alpha} \OfPrn{\bm{q}_i} \Bigr) } \, ,
\label{eq:Excitonic_SecondQuantization}
\end{equation}
where the electronic ONVs

\begin{equation}
\begin{aligned}
\ket{\bm{0}}        &= \ket{\alpha_0^{(1)} \cdots \alpha_0^{(N_\text{mol})}} \\
\ket{\bm{\alpha}_i} &= \ket{\alpha_0^{(1)} \cdots  \alpha_1^{(i)} \cdots \alpha_0^{(N_\text{mol})}} \\
\end{aligned}
\label{eq:VibronicDefinition}
\end{equation}
are given in the representation reported of Eq.~(\ref{eq:VibronicWavefunctionParametrization2}). The normal modes of the $i$-th monomer are labeled as $\bm{q}_i$, $\mathcal{V}_i^\text{GS}$ is the ground-state PES of the $i$-th monomer and $\mathcal{V}_i^{\alpha}$ is its $\alpha$-th excited state PESs. Eq.~(\ref{eq:Excitonic_SecondQuantization}) can be further simplified by approximating $\mathcal{V}_i^\text{GS}$ as a harmonic surface

\begin{equation}
\mathcal{V}_i^\text{GS} \OfPrn{\bm{q}_i} \approx \frac{1}{2} \sum_{j=1}^{N_\text{vib}} \omega_j^2 q_{i,j}^2 \, .
\label{eq:HarmonicPES}
\end{equation} 

We further assume that the harmonic frequencies $\omega_j$ and the normal modes are the same for all the monomers in the aggregate ($q_{i,j}=q_j$). Finally, we approximate $\mathcal{V}_i^{\alpha} \OfPrn{\bm{q}_i}$ with the linear coupling model,

\begin{equation}
\mathcal{V}_i^{\alpha} \OfPrn{\bm{q}_i} \approx \mathcal{V}_i^\text{GS} \OfPrn{\bm{q}_i} 
+ \sum_{j=1}^{N_\text{vib}} k_{\alpha,j} q_{i,j}
\label{eq:LCM_approximation}
\end{equation}
$k_{\alpha,j}$ being the $j$-th component of the gradient of the PES associated to the $\alpha$-th electronically excited state. Following the protocol proposed in Ref.~\citenum{Ivanov2015_ExcitonicReview}, we included in Eqs.~(\ref{eq:HarmonicPES}) and (\ref{eq:LCM_approximation}) only 10 modes associated to the largest value of the gradient. To include excitation transfer effects, a coupling term $\mathcal{V}_\text{c}$ is introduced

\begin{equation}
\mathcal{V}_\text{c} = \sum_{\langle\alpha,\alpha'\rangle} \ket{\bm{\alpha}} \bra{\bm{\alpha'}} \, J_{\alpha,\alpha'}
\label{eq:OffDiagonal_Excitonic}
\end{equation}
where the sum $\langle\alpha,\alpha'\rangle$ includes only neighboring terms. 

A second quantization form of Eqs.~(\ref{eq:HarmonicPES}) and (\ref{eq:LCM_approximation}) is trivially obtained from the relations given in Eq.~(\ref{eq:CanonicalQuantization}) for the nuclear part and Eq.~(\ref{eq:ElectronicSecondQuantization}) for the electronic one. We sorted the sites on the DMRG lattice as in Ref.~\citenum{Ren2018_TDDMRG-Temperature} so that, by employing the notation introduced above, the sorted ONV is $\ket{\alpha_0^{(1)} \alpha_1^{(1)} n_1^{(1)} \cdots \alpha_0^{(2)} \alpha_1^{(2)} n_1^{(2)} \ldots}$. Hence, the sites are sorted based on the monomer on which they are centered. Within each monomer, the sites associated to an electronically excited state appear before the nuclear ones.

\begin{figure}[htbp!]
	\centering
	\includegraphics[width=.75\textwidth]{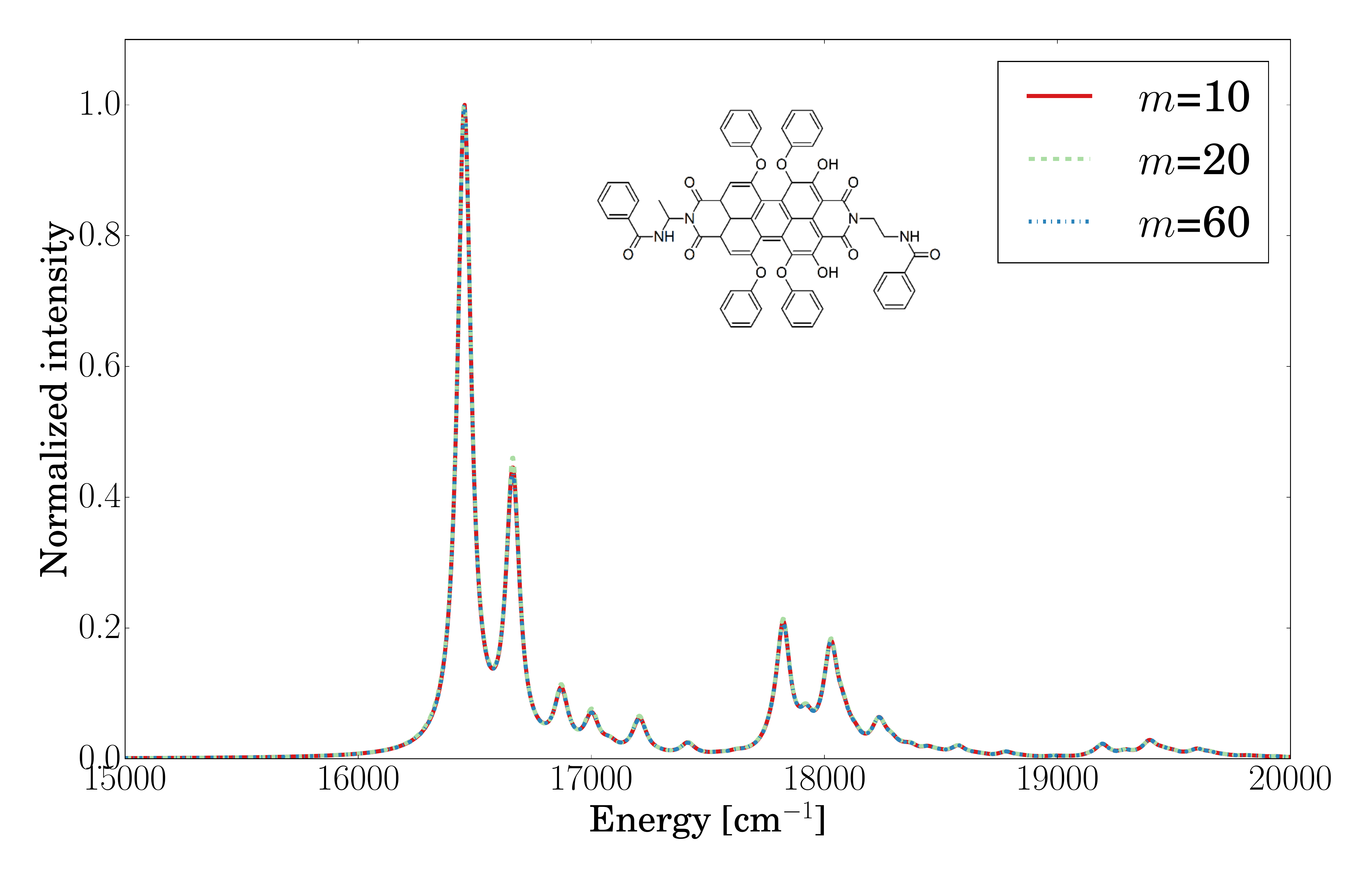}
	\caption{Absorption spectrum of the PBI-I monomer (Lewis structure reported in the spectrum) calculated with TD-DMRG for different values for the bond dimension $m$. All calculations were performed with $N_\text{max}=10$ and a time-step of 1 fs for an overall propagation time of 4000 fs.}
	\label{fig:Aggregate_1Molecule}
\end{figure}

\begin{figure}[htbp!]
	\centering
	\includegraphics[width=.75\textwidth]{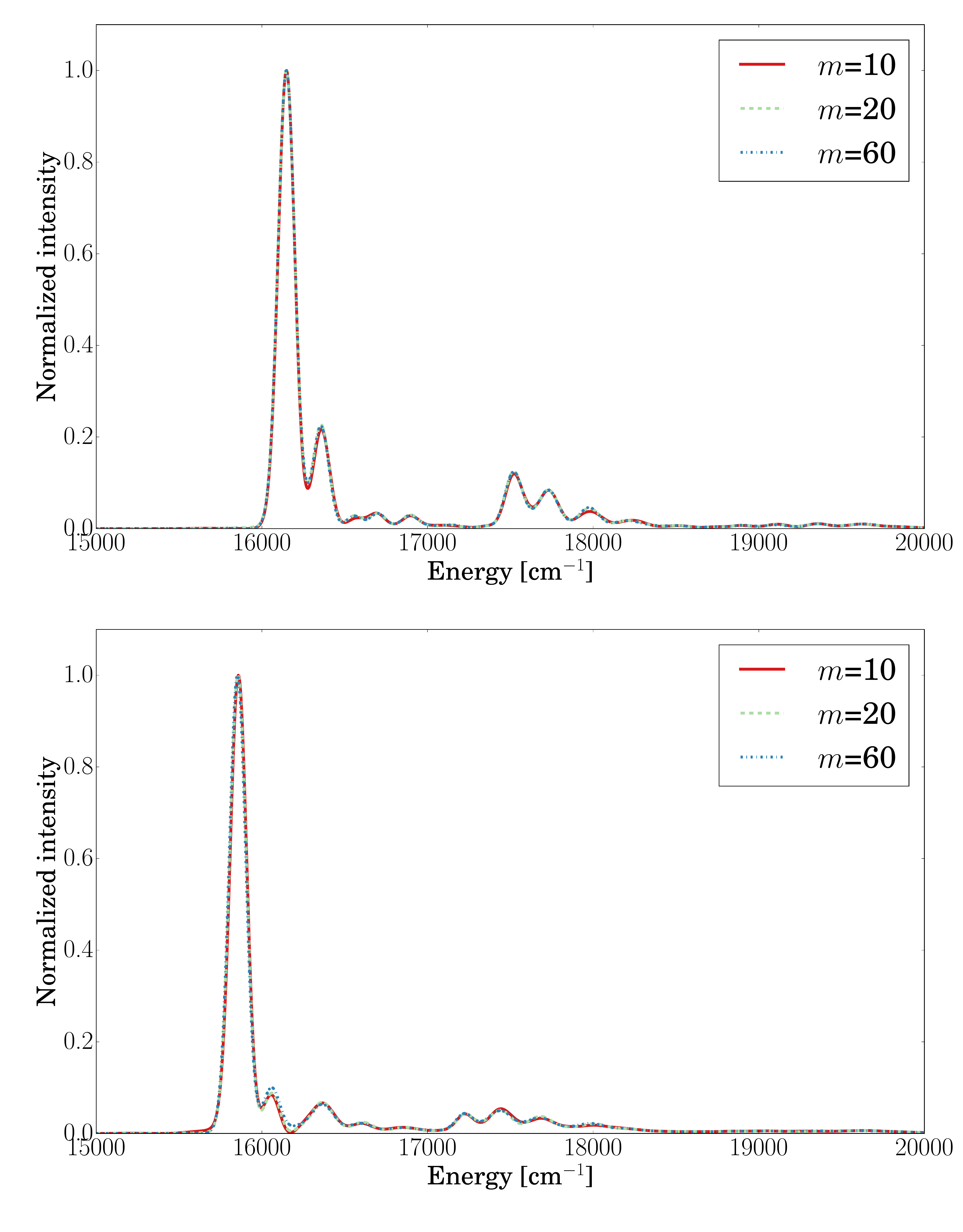}
	\caption{Absorption spectrum of the PBI-I dimer (upper panel) and hexamer (lower panel) calculated with TD-DMRG for different values for the bond dimension $m$. All calculations were performed with $N_\text{max}=10$ and a time-step of 1 fs for an overall propagation time of 4000 fs.}
	\label{fig:Aggregate_2Molecule}
\end{figure}

We first assess the stability of our MPS/MPO formulation of TD-DMRG with a single-site optimizer with respect to the integration time step $\Delta t$ and the bond dimension $m$. Figure~\ref{fig:Aggregate_1Molecule} reports the absorption spectrum of a single PBI-I molecule obtained with $m$=10, 20, and 60, a time step of 1 fs and an overall propagation time of 4000 fs. The absorption spectrum is fully converged already for $m$=10 and increasing $m$ to 20 or 60 does not modify the overall bandshape. The same trend is observed also for the dimer, as reported in the upper panel of Figure~\ref{fig:Aggregate_2Molecule}, where a fully converged spectrum is obtained with $m$=20. We find that, for the dimer, the initial state of the propagation is a coherent superposition of excitations localized on each monomer. Also for the hexamer, which produces a DMRG lattice with 66 sites, a fully converged spectrum is obtained with $m$=20. This indicates that the convergence of TD-DMRG can be largely independent on the system size. This is not surprising, since the Holstein Hamiltonian given in Eq.~(\ref{eq:Excitonic_SecondQuantization}) contains long-range interactions only between electronic degrees of freedom, while no long-range interactions are present for the vibrational part. These results suggest that the same holds true also for the time-dependent wavepacket during the whole propagation. We note that the $J$-coupling is classified as long-range because the sites are sorted in the lattice such that electronically excited states are not placed close one to the other. As will be discussed in more detail below, other sortings can be devised in which the $J$-coupling is short-ranged and that lead to a different convergence rate of DMRG.

\begin{figure}[htbp!]
	\centering
	\includegraphics[width=.75\textwidth]{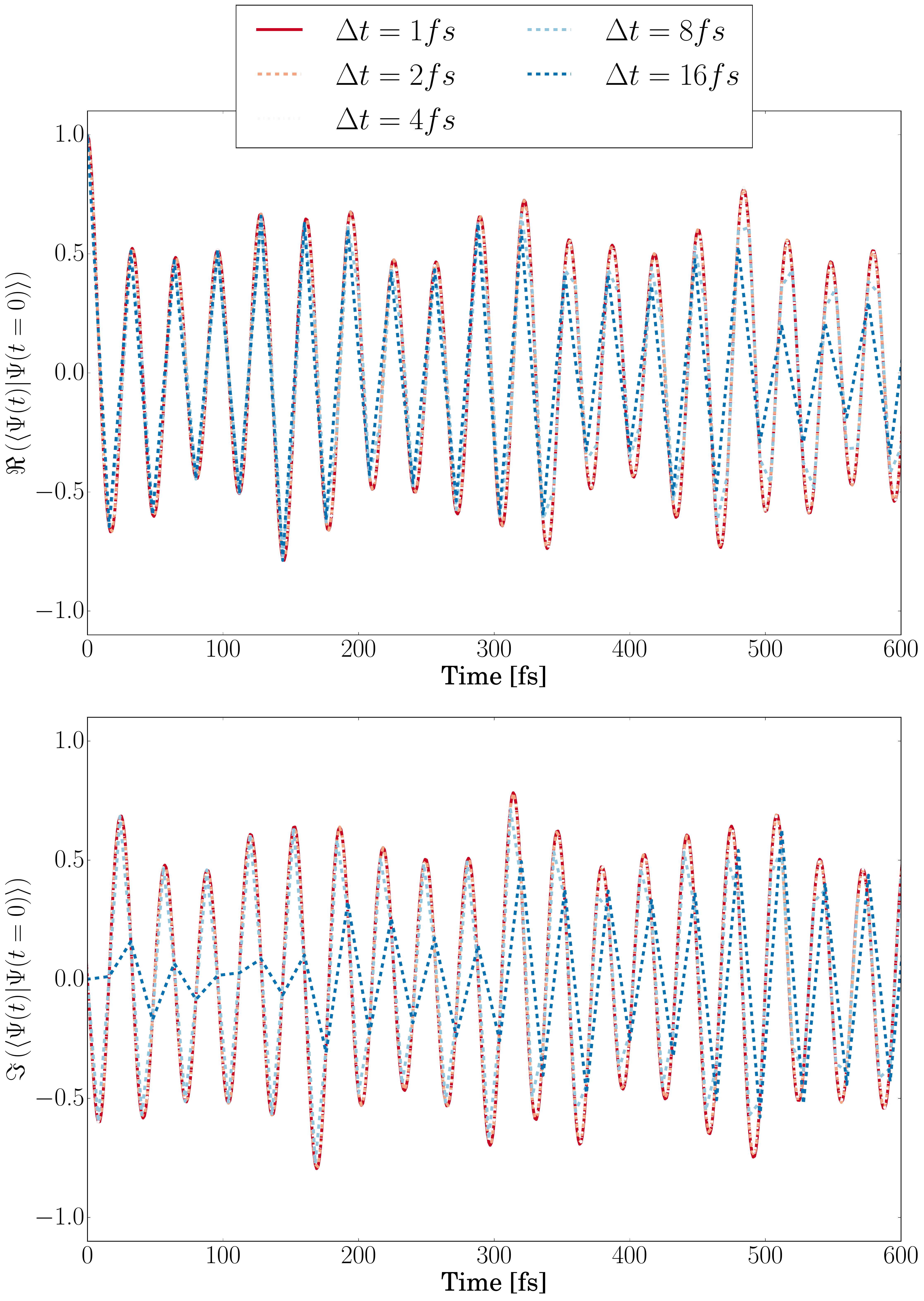}
	\caption{Time-dependent autocorrelation function of the total vibronic wave function the PBI-I dimer calculated with TD-DMRG for different values of the time step $\Delta t$ employed in the propagation. All calculations were performed with $N_\text{max}=10$ and $m$=20.}
	\label{fig:Aggregate_TimeStep}
\end{figure}

For a given value of $m$, the accuracy of TD-DMRG depends on the step size $\Delta t$ chosen to discretize the propagation. The Lie-Trotter factorization of Eq.~(\ref{eq:LieTrotter}) is valid in the limit of short time steps. Therefore the accuracy is expected to deteriorate with increasing $\Delta t$. The error introduced in the Lie-Trotter factorization is quadratic in the time step $\Delta t$, proportional to the commutator of the local operators and can, in principle, be estimated.\cite{Verstraete2013_GeometryMPS,Schollwoeck2019_TDDMRG-Review} Here, we only compare the spectra obtained with different time steps. Figure~\ref{fig:Aggregate_TimeStep} reports the real (upper panel) and imaginary (lower panel) part of the time-dependent autocorrelation function of the PBI-I dimer for values of the time-step between 1 and 16 fs. For the first 600 fs of the propagation, the autocorrelation function is the same independent of the time step, suggesting that the Lie-Trotter approximation is reliable and time steps that are significantly larger than the ones employed in other MCTDH and TD-DMRG studies on this system can be safely employed.\cite{Kuhn2012_ExcitonicPerylene-6Mode,Ren2018_TDDMRG-Temperature} We note that the time step affects the numerical Fourier transform of the autocorrelation function. The maximum frequency that can be detected with a numerical Fourier transform is $1/\Delta t$,\cite{Shannon1949_FFT} and therefore, a large value for the time step impedes the calculation of the absorption spectrum in a wide energy range.

\begin{figure}[htbp!]
	\centering
	\includegraphics[width=.75\textwidth]{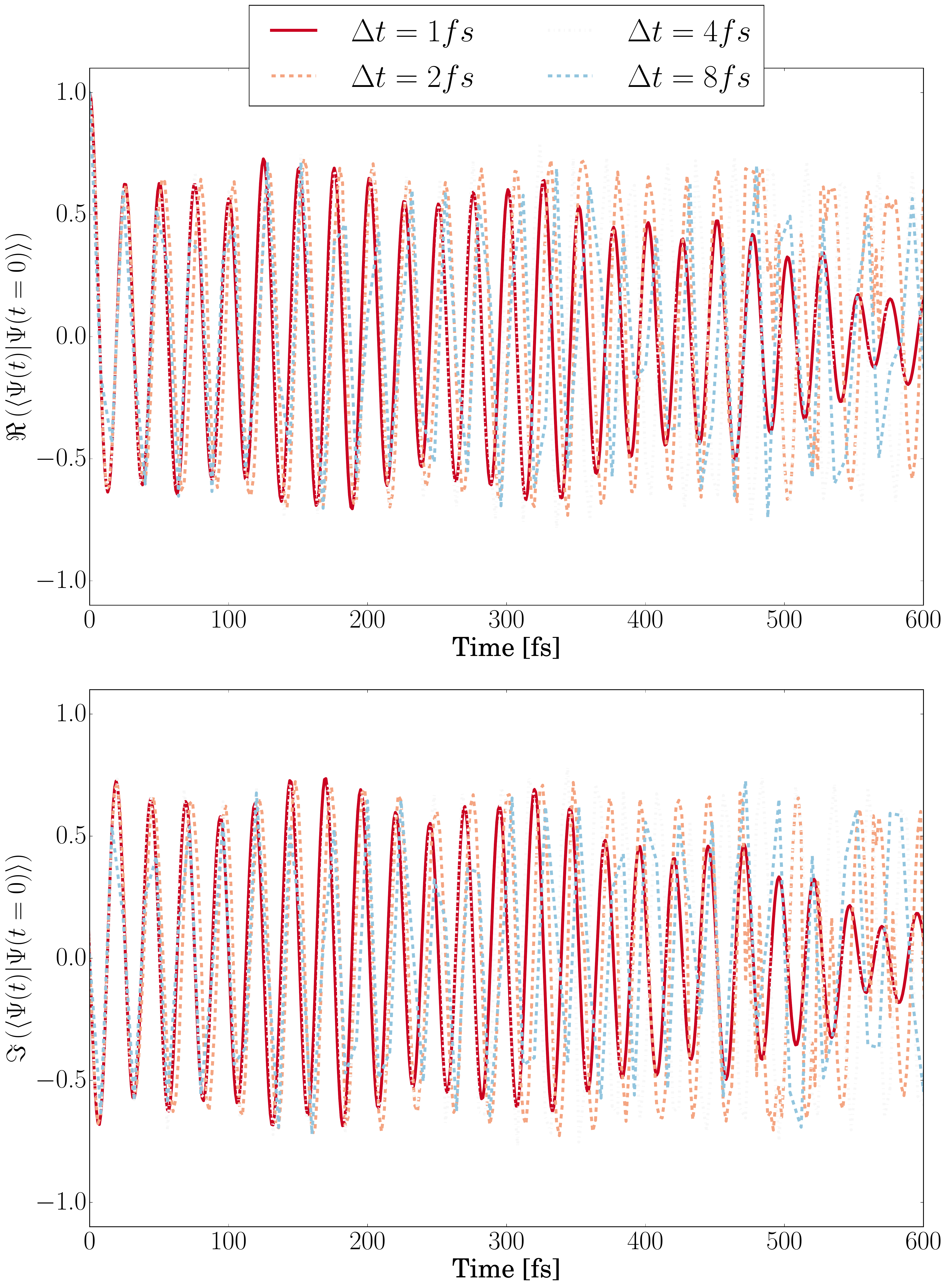}
	\caption{Time-dependent autocorrelation function the PBI-I hexamer calculated with TD-DMRG for different values of the time step $\Delta t$ employed in the propagation. All calculations were performed with $N_\text{max}=10$ and $m$=20.}
	\label{fig:Aggregate_TimeStep_6Molecule}
\end{figure}

The graphical representation of the autocorrelation function for the hexamer obtained with different time steps is reported in Figure~\ref{fig:Aggregate_TimeStep_6Molecule}. We note that larger differences between the results obtained with different time steps are observed in this case. The autocorrelation function obtained with $\Delta t$=8~fs agrees with the one obtained with $\Delta t$=1~fs only up to 300~fs, and then they diverge. The error introduced in the Lie-Trotter factorization is proportional to the sum of all the possible commutators between the terms into which the Hamiltonian is partitioned. The number of these terms increases with system size, and therefore, a smaller time step is expected to be required for larger DMRG lattices.

\begin{figure}[htbp!]
	\centering
	\includegraphics[width=.75\textwidth]{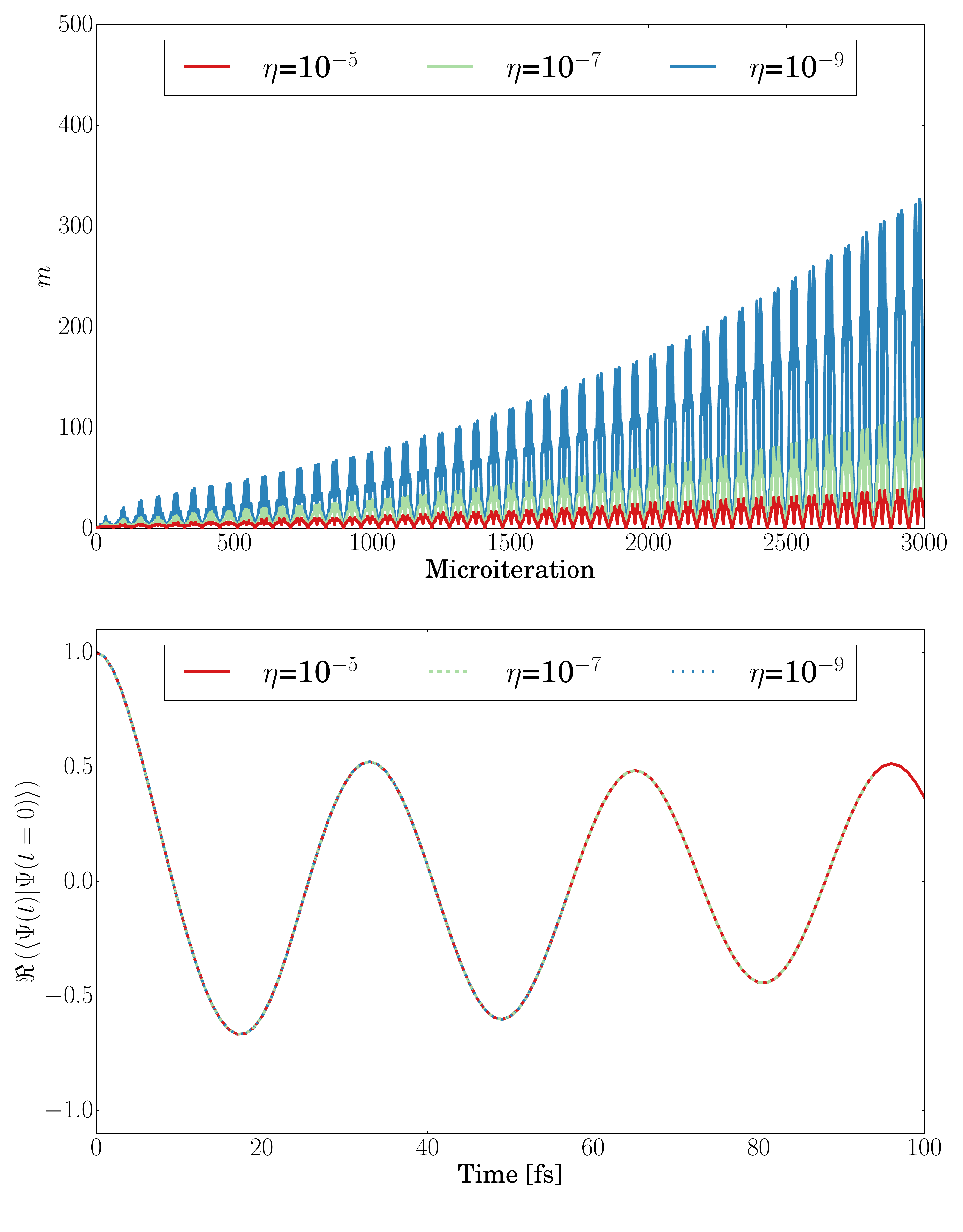}
	\caption{Upper panel: bond dimension $m$ obtained with the dynamical block state selection algorithm and the two-site TD-DMRG for different values for the threshold on the truncation error $\eta$ as a function of the microiteration number. Lower panel: real part of the autocorrelation function obtained with two-sites TD-DMRG for different values of the threshold on the truncation error $\eta$ as a function of the propagation time $t$.}
	\label{fig:Aggregate_TwoMolecule_DBSS}
\end{figure}

We now investigate the reliability of the two-site propagation scheme for the dimer of PBI-I. As already mentioned in the theoretical section, the main advantage of a two-site TD-DMRG algorithm over a single-site one is the possibility of adapting dynamically the bond dimension based on the accuracy that is required in the representation of the wavefunction. We calculate the absorption spectrum of the dimer using different values for the threshold $\eta$ on the truncation error. We report in the upper panel of Figure~\ref{fig:Aggregate_TwoMolecule_DBSS} a graphical representation of the bond dimension $m$ obtained with two-site TD-DMRG for different values of $\eta$ ranging from 10$^{-5}$ to 10$^{-9}$. For each value of $\eta$, the bond dimension $m$ increases continuously with time. As highlighted in Ref.~\citenum{Pollman2012_UnboundedEntanglement-TD}, the entanglement of a quantum system grows during a non-equilibrium time evolution, and therefore, the bond dimension $m$ required to represent the wavefunction with a given accuracy (that is proportional to $\eta$) increases as well. Our calculations suggests that this is also the case for the PBI-I dimer. We note that, even if a steep increase for $m$ is observed by reducing $\eta$ from 10$^{-5}$ to 10$^{-9}$, the choice of $\eta$ will have a negligible effect on the autocorrelation function (lower panel of Figure~\ref{fig:Aggregate_TwoMolecule_DBSS}). This suggests that a rather high value of $\eta$ can be sufficient to obtain a fully-converged autocorrelation function, even if obtained for non-converged wavefunctions.

We recall that the autocorrelation function was obtained from the overlap of the time-dependent wavefunction $\ket{\Psi(t)}$ with the initial state $\ket{\Psi(0)}$. The latter state is a product state that can be represented with an MPS with $m$=1. It is, therefore, possible that the overlap with such a compact MPS converges faster than the representation of the full-dimensional wavefunction. This hypothesis agrees with the analysis reported in Ref.~\citenum{Kloss2018}, where it was shown that even if, for fixed $m$, the truncation error grows with increasing time in single-site TD-DMRG, the calculation of averaged quantities is stable.

\begin{figure}[htbp!]
	\centering
	\includegraphics[width=.75\textwidth]{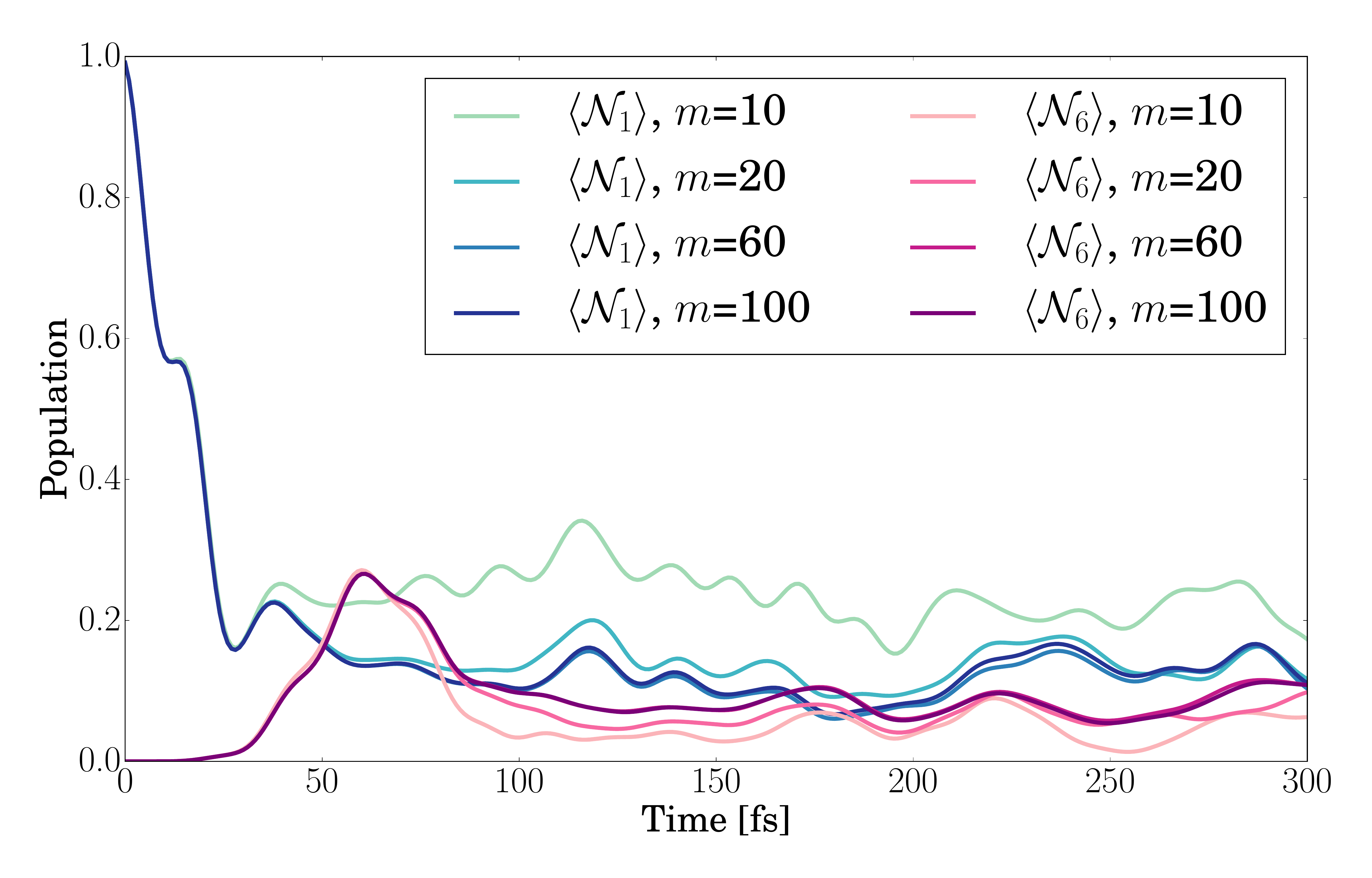}
	\caption{Electronic population of the first electronically excited state of the first (blue lines) and last (red lines) molecule of the PBI-I hexamer obtained with single-site DMRG, $N_\text{max}=10$, a time-step of 1 fs and different values for the bond dimension $m$.}
	\label{fig:Aggregate_6Molecule_PopFirstsLast}
\end{figure}

Besides the autocorrelation function, the population of each excited state as a function of time is easily extracted from TD-DMRG propagations through the average of the number operator $\mathcal{N}_\alpha$, defined as

\begin{equation}
\mathcal{N}_\alpha = a_{\alpha}^+ a_{\alpha} \, .
\label{eq:NumberOperator}
\end{equation}

The number operator is local and can be easily encoded as an MPO. The dynamics of the population of excited states is particularly interesting if, as an initial state, we consider an excitation localized on one terminal molecule of the aggregate. In this case, tracking the average of $\mathcal{N}_\alpha$ enables one to calculate the rate of the transfer of excitation along the aggregate. We show in Figure~\ref{fig:Aggregate_6Molecule_PopFirstsLast} the population of the two terminal monomers of the PBI-I hexamer, $\langle \mathcal{N}_1 \rangle$ and $\langle \mathcal{N}_6 \rangle$, as a function of time $t$. The first discrepancies between the results obtained for different values of $m$ are observed after 25~fs, the population of the first site being overestimated with $m$=10. The populations obtained with with $m$=20 and 60 agree qualitatively and the results obtained with $m$=60 are identical to the $m$=100 ones.

\begin{figure}[htbp!]
	\centering
	\includegraphics[width=.75\textwidth]{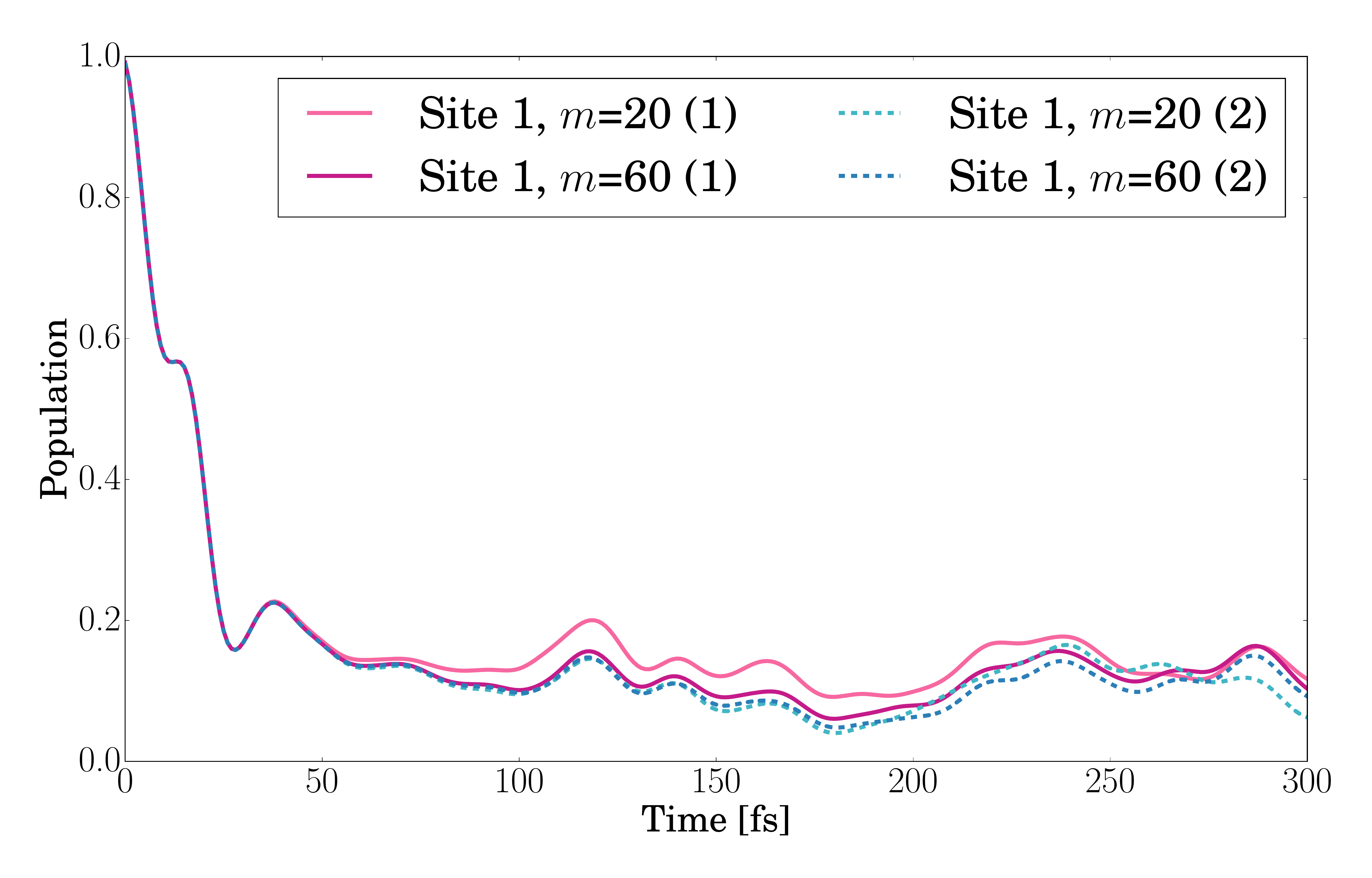}
	\caption{Electronic population of the first electronically excited state of the first molecule of the PBI-I hexamer obtained with DMRG by employing two different sortings (1: electronic sites placed close to the vibrational one, 2: electronic sites placed one close to each other.). Calculations were performed with $N_\text{max}=10$, a time-step of 1 fs, and different values for the bond dimension $m$.}
	\label{fig:Aggregate_6Molecule_PopFirstsLast_EleFirst}
\end{figure}

The ordering of the DMRG lattice given in Ref.~\citenum{Ren2018_TDDMRG-Temperature}, which we have employed so far, is not unique and can be expected to affect the convergence of TD-DMRG. To assess the relevance of the ordering, we analyze the population dynamics with an alternative ordering obtained as $\ket{\alpha_0^{(1)} \alpha_1^{(1)} \alpha_0^{(2)} \alpha_1^{(2)} n_1^{(1)} \cdots n_1^{(2)} \cdots}$. In practice, all the electronic degrees of freedom are placed before the vibrational ones. In the Holstein Hamiltonian, two interaction terms are present, the hopping term, given in Eq.~(\ref{eq:OffDiagonal_Excitonic}), and the linear vibronic term of Eq.~(\ref{eq:LCM_approximation}). The relative strengths of these coupling terms affects the efficiency of the two orderings. We show in Figure~\ref{fig:Aggregate_6Molecule_PopFirstsLast_EleFirst} the population of the excited monomer, at one end of the aggregate, as a function of time for the two orderings described above. If the electronic sites were put close to each other, converged results had been obtained with $m$=20 up to 200 fs whereas, with the other ordering, deviations had been observed already at 100 fs. This suggests that the $J$ coupling is, in this specific case, predominant and that the electronic sites are more strongly entangled than the vibrational ones. Clearly, a quantitative analysis of the entanglement would require the calculation of the von Neumann entropy.\cite{Legeza2003_OrderingOptimization,Rissler2006_QuantumInformationOrbitals,Boguslawski2012_OrbitalEntanglement} However, such an analysis is beyond the scope of the present work.

\subsection{Excited-state dynamics of pyrazine}

We now calculate with TD-DMRG the low-energy vibrationally-resolved electronic spectrum of pyrazine for the S$_2$ $\leftarrow$ S$_0$ excitation. The S$_2$ excited state of pyrazine is strongly coupled to S$_1$ by non-adiabatic couplings, and therefore, an accurate absorption spectrum is obtained only with a vibronic Hamiltonian treating simultaneously these two states. For this work, we chose the vibronic Hamiltonian derived in Ref.~\citenum{Woywod1994_PyrazinePES} with the updated parameters reported in Ref.~\citenum{Raab1999_PyrazineDiabatic},

\begin{equation}
\mathcal{H} = \left( 
{\begin{array}{cc} 
	\mathcal{T}_{S_1}(\bm{q}) &             0             \\
	0               & \mathcal{T}_{S_1}(\bm{q}) \\
	\end{array} } \right) + \left( 
{\begin{array}{cc} 
	\mathcal{V}_{S_1}(\bm{q}) & \mathcal{V}_{12}(\bm{q}) \\
	\mathcal{V}_{12}(\bm{q})  & \mathcal{V}_{S_2}(\bm{q}) \\
	\end{array} } \right) \, ,
\label{eq:PyrazineVibrationalHamiltonian}
\end{equation}
where $\mathcal{V}_{S_1}(\bm{q})$ and $\mathcal{V}_{S_2}(\bm{q})$ are the potential energy surfaces of the S$_1$ and S$_2$ electronic states, respectively, and are obtained as a second-order Taylor expansion around the equilibrium geometry of the ground state. The off-diagonal coupling $\mathcal{V}_{12}(\bm{q})$ is also expressed as a second-order Taylor expansion. The Hamiltonian of Eq.~(\ref{eq:PyrazineVibrationalHamiltonian}) has become a standard benchmark system to probe the efficiency of new semiclassical\cite{Thoss2000_PyrazineSemiclassical,Wang2001_Pyrazine-IVR,Zimmermann2014_Pyrazine-Semiclassical} and quantum\cite{Raab1999_PyrazineDiabatic,Burghardt2008_VibronicMCTDH,Vendrell2011_ML-MCTDH,Lasorne2015_vMCG-Review,Saller2017_Adaptive-QuantumDynamics} dynamics algorithms. A full-dimensional quantum simulation of the dynamics of pyrazine, including all 24 vibrational degrees of freedom treated as independent and at the same level of accuracy is prohibitive for most quantum dynamics algorithms available in the literature, and reduced-dimensionality schemes must be adopted.\cite{Raab1999_PyrazineDiabatic,Burghardt2008_VibronicMCTDH,Saller2017_Adaptive-QuantumDynamics} Alternatively, the computational costs can be reduced by adopting ML-MCTDH.\cite{Vendrell2011_ML-MCTDH} In the present work, we demonstrate how TD-DMRG can be applied to the simulation of the full-dimensional quantum dynamics of pyrazine without the need of introducing any hierarchical treatment of the degrees of freedom.

\begin{figure}[htbp!]
	\centering
	\includegraphics[width=.75\textwidth]{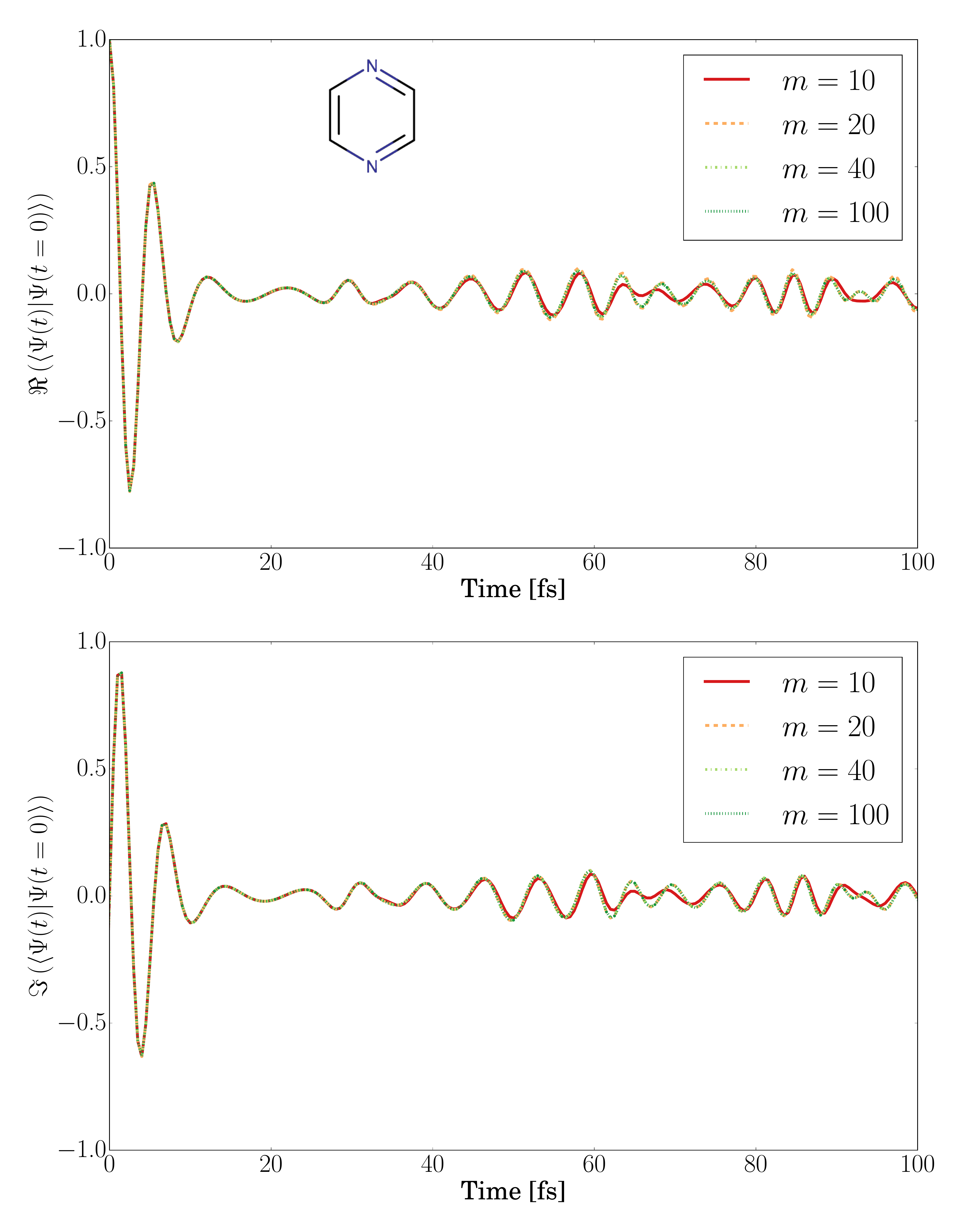}
	\caption{Real (upper panel) and imaginary (lower panel) part of the autocorrelation function of the total vibronic wave function of pyrazine (Lewis structure reported in the upper panel) after photoexcitation to the S$_2$ electronic state as a function of time for different values for the bond dimension $m$. Calculations were performed with $N_\text{max}=10$ and a time-step of 0.5 fs.}
	\label{fig:Pyrazine4ModeAutocorr}
\end{figure}

We begin our analysis with the simulation of the dynamics of pyrazine upon photoexcitation for a reduced-dimensionality model including only four modes, namely $\nu_{10a}$, $\nu_{6a}$, $\nu_1$, and $\nu_{9a}$ (following the notation of Ref.~\citenum{Raab1999_PyrazineDiabatic}), which are the most relevant modes governing the dynamics. In the occupation number vector representation, the initial state of the propagation is $\ket{0_{S_1} 1_{S_2},0000}$ since, under the Franck-Condon approximation, the molecule is excited to the S$_2$ electronic state without modifying its vibrational wavefunction.

\begin{figure}[htbp!]
	\centering
	\includegraphics[width=.75\textwidth]{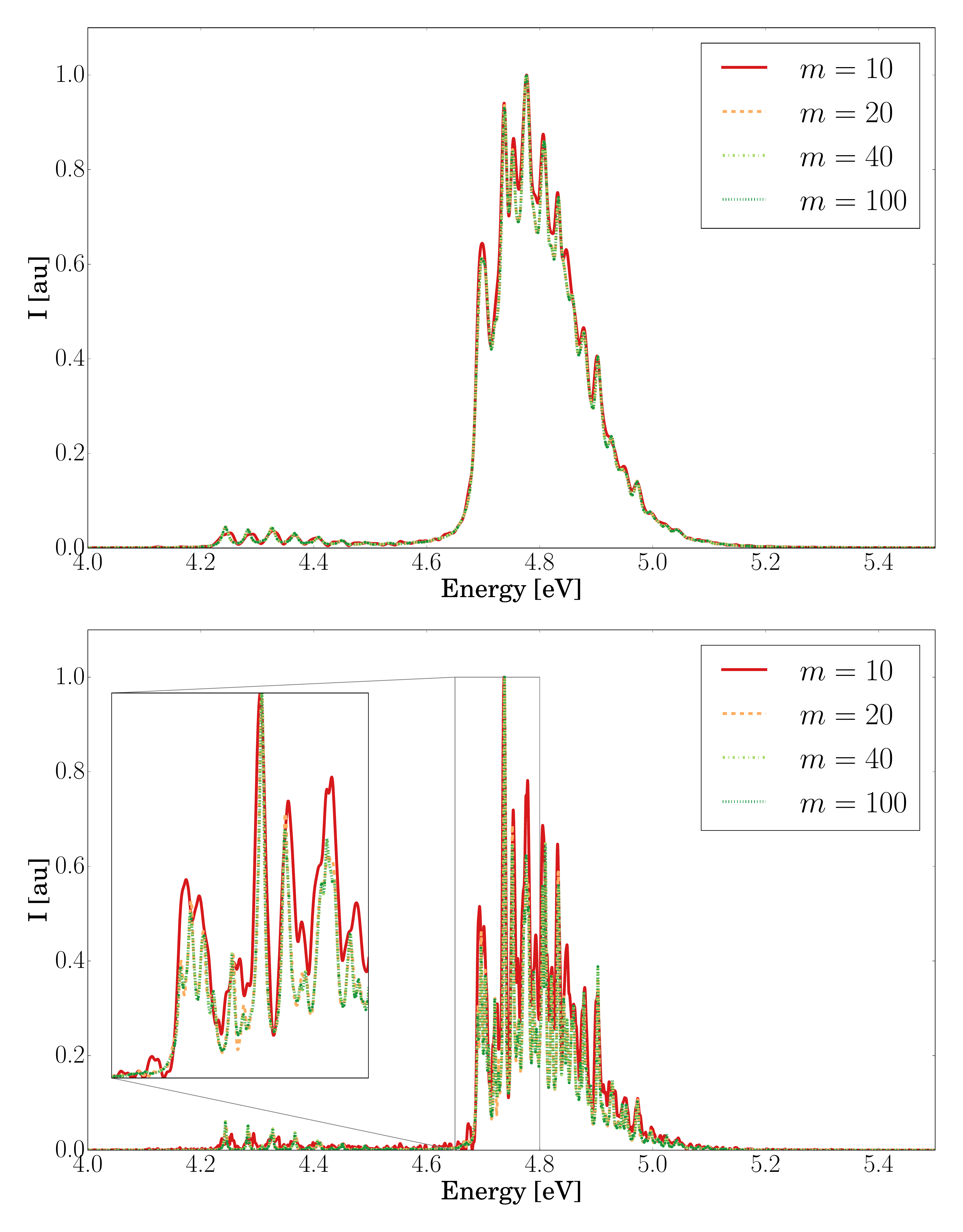}
	\caption{Vibrationally-resolved vibronic spectrum of pyrazine for the S$_2$ $\leftarrow$ S$_0$ transition calculated with single-site TD-DMRG with different values for the bond dimension $m$. Calculations were performed with $N_\text{max}=10$, a time-step of 0.5 fs, and an overall propagation time of 4000 fs. Broadening effects were reproduced with an exponential damping function with $\tau$=50~fs$^{-1}$ (upper panel) and 200~fs$^{-1}$ (lower panel).}
	\label{fig:Pyrazine4Mode_VibronicSpectrum}
\end{figure}

We report in Figure~\ref{fig:Pyrazine4ModeAutocorr} the real and imaginary part of the autocorrelation function $C(t)=\OvI{\Psi(t)}{\Psi(0)}$ obtained by TD-DMRG propagation with different values of $m$. In all calculations we chose $N_\text{max}=10$ and a time step $\Delta t$ of 0.5~fs. For the first 50 fs of the propagation, the autocorrelation function converged already with $m$=10 with negligible variations observed by increasing $m$ up to 100. Starting from $t=50$ fs, small deviations are observed and only with $m$=20 converged results are obtained. We report in Figure~\ref{fig:Pyrazine4Mode_VibronicSpectrum} the vibronic spectrum obtained from these propagations. We include broadening effects with an exponential damping function with $\tau$=50~fs$^{-1}$ and 200~fs$^{-1}$. For the lowest-resolution spectrum ($\tau$=50~fs$^{-1}$) significant discrepancies are visible between the spectra obtained with $m$=10 and higher values of $m$. By increasing the resolution ($\tau$=200~fs$^{-1}$), slight discrepancies between the spectra are observed also between the $m$=20 and $m$=40 spectra. However, no additional changes emerge by further increasing $m$.

\begin{figure}[htbp!]
	\centering
	\includegraphics[width=.75\textwidth]{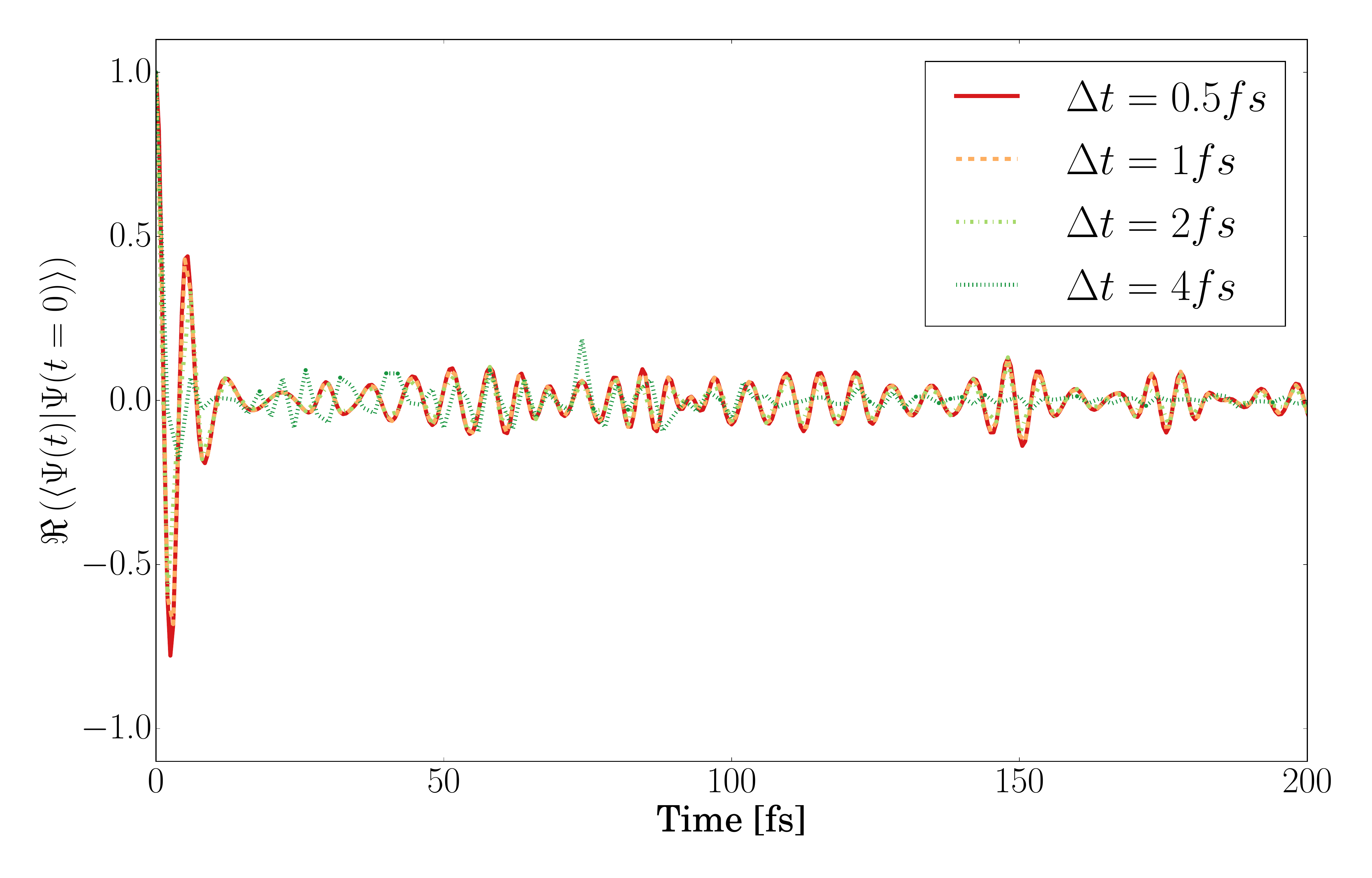}
	\caption{Real part of the autocorrelation function of pyrazine after photoexcitation to the S$_2$ electronic state as a function of the time calculated with single-site TD-DMRG with different values for the integration time-step $\Delta t$. Calculations were performed with $N_\text{max}=10$ and $m$=20.}
	\label{fig:PyrazineTimeSteps}
\end{figure}

In Figure~\ref{fig:PyrazineTimeSteps}, we compare the real part of the time-dependent autocorrelation function obtained for $m$=20 and different values for the integration time step. For the first 200 fs of the propagation, we note that the integration time can be increased up to 2 fs without a significant loss of accuracy, whereas strong deviations are observed by setting $\Delta t$ to 4 fs. Hence, a smaller time step is sufficient compared to the one needed to reach convergence for the previous example. This could be due to the fact that the Hamiltonian contains both short- and long-range interactions, and therefore, its MPO representation requires a higher bond dimension to maintain accuracy. Applying this MPO on an MPS increases the bond dimension faster than in the previous example, and therefore, a smaller time step is needed to ensure that the wavefunction remains on the manifold of the MPSs with a given value of $m$. Nevertheless, the time step is still higher than the one employed in other TD-DMRG formulations.\cite{Ren2018_TDDMRG-Temperature}

\begin{figure}[htbp!]
	\centering
	\includegraphics[width=.75\textwidth]{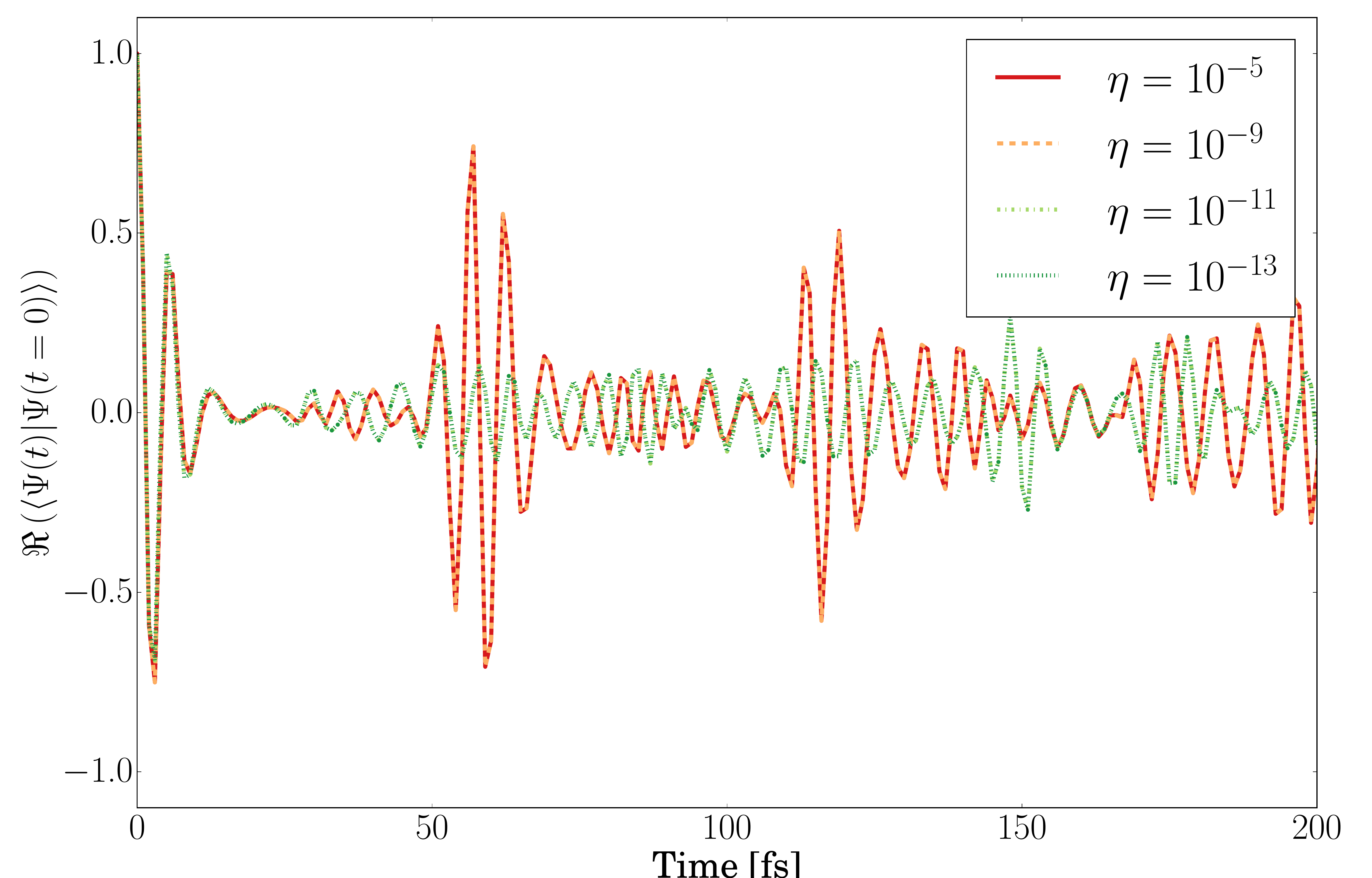}
	\caption{Real part of the autocorrelation function of pyrazine after photoexcitation to the S$_2$ electronic state as a function of the time calculated with two-sites TD-DMRG with different values for the  truncation error $\eta$. Calculations were performed with $N_\text{max}=10$ and $m$=20.}
	\label{fig:PyrazineTwoSites}
\end{figure}

We conclude our analysis of this 4-mode reduced-dimensionality Hamiltonian of pyrazine by investigating the two-site integration scheme. Figure~\ref{fig:PyrazineTwoSites} shows the graphical representation of the real part of the time-dependent autocorrelation function obtained with $N_\text{max}$=10 and different values of the truncation error $\eta$. With values of $\eta$ larger than 10$^{-10}$, the decay of the autocorrelation function is slower than with $\eta$ smaller than 10$^{-10}$, with constructive interference approximately every 50 fs. Comparing the results reported in Figure~\ref{fig:PyrazineTwoSites} with Figure~\ref{fig:Pyrazine4ModeAutocorr} we note that a truncation error of $10^{-11}$ is sufficient to reach converged results.

\begin{figure}[htbp!]
	\centering
	\includegraphics[width=.75\textwidth]{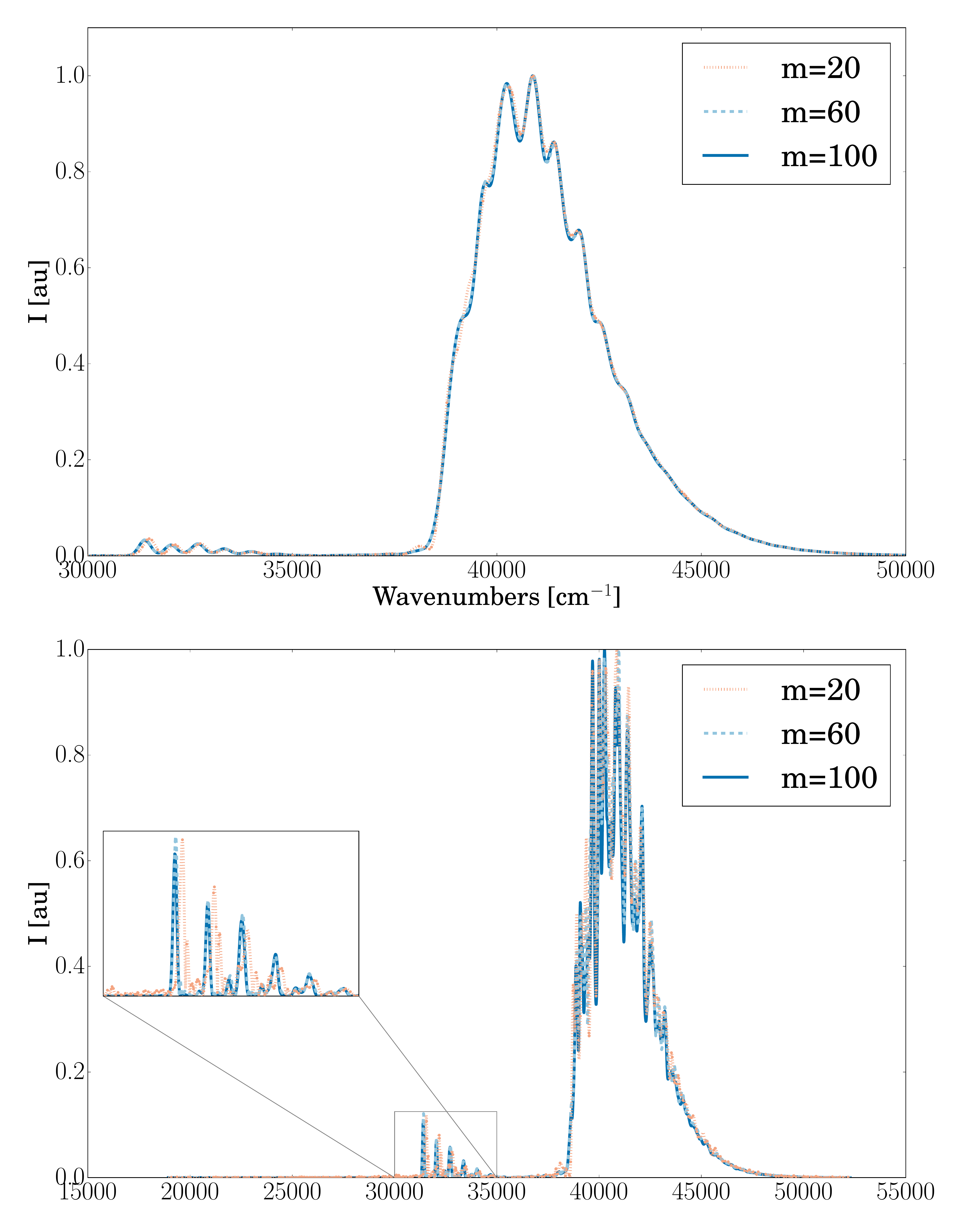}
	\caption{Upper panel: Absorption spectrum of pyrazine calculated on the 24-mode, full-dimensional system with single-site TD-DMRG for different values for the bond dimension $m$. Calculations were performed with $N_\text{max}=10$ and a damping factor $\tau$ of 50 fs$^{-1}$ (upper panel) and 200 fs$^{-1}$ (lower panel) was employed to model broadening effects.}
	\label{fig:PyrazineFullDim}
\end{figure}

Considering the absorption spectrum of pyrazine based on the 24-modes full-dimensional vibronic Hamiltonian, we report in the upper panel of Figure~\ref{fig:PyrazineFullDim} the vibronic spectrum obtained with different values for the bond dimension $m$ employing a damping factor $\tau$ of 50 fs$^{-1}$. We note that, despite the strong increase in system size (the DMRG lattice is now composed by 26 sites), the overall bandshape has already been qualitatively reproduced for 20 renormalized block states. By increasing $m$ to 40 and 60, only slight deviations are observed in the low-energy tail of the higher-energy band. By increasing the resolution of the spectrum through a damping factor of 200 fs$^{-1}$, larger differences are observed as can be seen in the lower panel of Figure~\ref{fig:PyrazineFullDim}. For the lower-energy band, an overall blue-shift is observed for $m$=20. Nevertheless, the spectrum obtained for $m$=60 is superimposable to that for $m$=100. A similar effect is observed for the more intense S$_2$ $\leftarrow$ S$_0$ band. Increasing $m$ to 60 corrects the band position, but does not provide a fully converged spectrum, since a significant redistribution of the different bands of the vibronic spectrum is observed by further increasing $m$ to 100. 

\subsection{Imaginary time propagation: vibrational states of C$_2$H$_4$}

The range of application of TD-DMRG can be extended by generalizing it to imaginary-time propagations. For example, the time evolution of thermal ensembles can be obtained within the so-called ancillla state approach by finite-time imaginary time evolution of a matrix product state.\cite{Vidal2004_MPO-DensityOperator,Schollwoeck2011_Review-DMRG,Karrash2012_FiniteTemperature-SpinChain} This approach, which was been generalized to an MPS/MPO formulation,\cite{Barthel2013_FiniteTemperature-MPO} was applied to the calculation of finite-temperature spectra of the aggregate studied at the beginning of this section.\cite{Ren2018_TDDMRG-Temperature} Imaginary-time propagations are also an alternative route to the sweep-based scheme to optimize MPSs. In fact, the following property holds:

\begin{equation}
\lim_{t \rightarrow \infty} e^{-t\mathcal{H}} \ket{\Psi_\text{MPS}} = \ket{\Psi_0} \, ,
\label{eq:ImaginaryTimePropagation}
\end{equation}
where $\ket{\Psi_\text{MPS}}$ is a random MPS and $\ket{\Psi_0}$ is the MPS representation of the ground state of the Hamiltonian $\mathcal{H}$. Eq.~(\ref{eq:ImaginaryTimePropagation}) is the basis of Diffusion Monte Carlo methods\cite{Lester1992_FixedNodeQMC} that solve the differential equation stochastically to avoid an uncontrolled increase in the basis size. In vibrational calculations, imaginary-time propagation of the MCTDH equations was applied to the calculation of the vibrational ground and excited states of several strongly anharmonic systems.\cite{Meyer2006_MCTDH-ImaginaryTime,Gatti2008_ImaginaryTime-Fluoroform,Manthe2012_Malonaldehyde-ImaginaryTime,Meyer2013_Malonaldehyde-ImaginaryTime} As for real-time propagations, the computational costs of MCTDH can be limited with TD-DMRG.\cite{Vidal2004_TEBD} Even if the sweep-based optimization is in most cases efficient enough, imaginary-time propagation can be easily generalized to excited states through,

\begin{equation}
\lim_{t \rightarrow \infty} e^{-\left(\mathcal{H} - \omega \right)^2 t^2} \ket{\Psi_\text{MPS}} = \ket{\Psi_\omega} \, ,
\label{eq:ImaginaryTimePropagation_ExcitedStates}
\end{equation}
where $\omega$ is an energy-shift parameter and $t$ is squared to match the units of the squared Hamiltonian. It can be easily shown\cite{AspuruGuzik2015_ImaginaryTime} that the limiting state $\ket{\Psi_\omega}$ is the excited state of $\mathcal{H}$ with energy closest to $\omega$. Time-evolution governed by the propagator $e^{-t^2 \left(\mathcal{H} - \omega \right)^2}$ can be simulated based on the algorithm described in Section~\ref{sec:theory} by employing the MPO representation of $\left(\mathcal{H} - \omega  \right)^2$ instead of the one of $\mathcal{H}$. As discussed in our previous work,\cite{Baiardi2019_HighEnergy-vDMRG} the calculation of matrix elements of powers of the Hamiltonian is more efficient and accurate within an MPS/MPO framework.

\begin{figure}[htbp!]
	\centering
	\includegraphics[width=.75\textwidth]{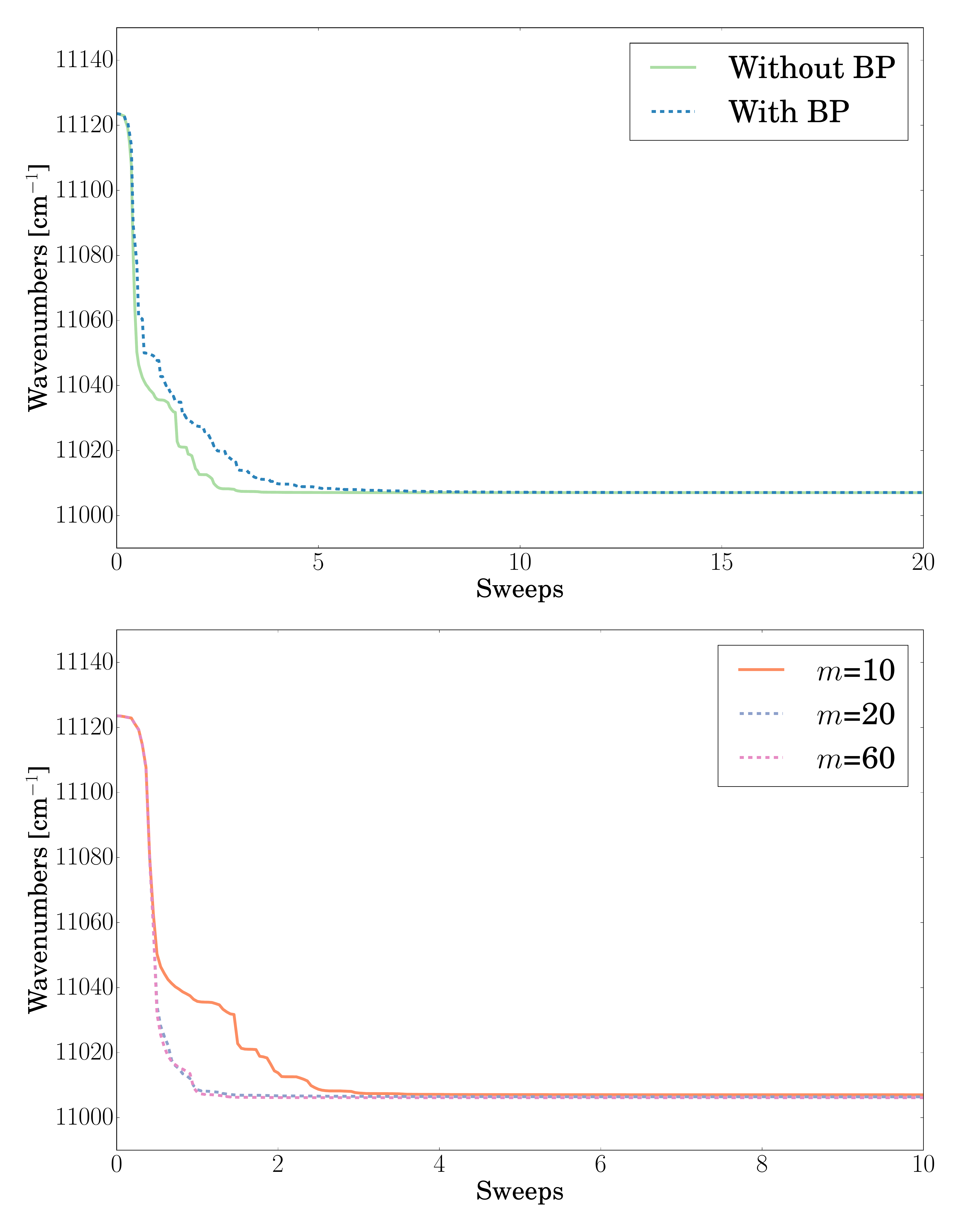}
	\caption{Upper panel: Energy of the vibrational ground state of ethylene as a function of the number of sweeps obtained with imaginary-time TD-DMRG. Calculations were performed with a time-step of 1~fs, N$_\text{max}$=6, and $m$=10. The time evolution was performed with the full propagation algorithm (dashed, blue line) and by not performing the back-propagation (BP) of the zero-site tensor (solid, green line). \\
		Lower panel: Energy of the vibrational ground state of ethylene as a function of the number of sweeps obtained with imaginary-time TD-DMRG. Calculations were performed with a time-step of 1~fs, N$_\text{max}$=6, and different values for the bond dimension $m$.}
	\label{fig:C2H4_ImaginaryTime}
\end{figure}

Here, we apply imaginary-time TD-DMRG to the optimization of the vibrational levels of ethylene, which we considered in previous work in the context of vDMRG.\cite{Baiardi2017_VDMRG,Baiardi2019_HighEnergy-vDMRG} We employ the same potential energy surface as in Ref.~\citenum{Baiardi2019_HighEnergy-vDMRG}, taken from Ref.~\citenum{Delahaye2014_EthylenePES} and converted to a fourth-order Taylor expansion in Cartesian normal modes by neglecting first-order $\pi$-$\pi$ coupling terms in the kinetic energy operator. The upper panel of Figure~\ref{fig:C2H4_ImaginaryTime} reports the results obtained with N$_\text{max}$=6 and $m$=10 for the non-shifted imaginary-time propagator obtained from Eq.~(\ref{eq:ImaginaryTimePropagation}). The energy converges within three sweeps, and therefore, the algorithm is as efficient as the standard time-independent optimization scheme.\cite{Baiardi2017_VDMRG} In Ref.~\citenum{Haegeman2016_MPO-TDDMRG}, it was noted that the back-propagation step of the zero-site tensor (the terms involving the $\mathcal{P}_i^{(2)}$ projector operators in Eq.~(\ref{eq:ProjectorExpression})) is not needed in imaginary-time propagations, in which the wavefunction for finite values of $t$ is not of interest and only the asymptotic limit of the propagation is needed. As reported in the upper panel of Figure~\ref{fig:C2H4_ImaginaryTime}, skipping the back-propagation step improves the rate of convergence of the energy, therefore confirming the previous observation. Based on these results, if not otherwise specified, imaginary-time evolutions are performed without back-propagating the zero-site tensor in the following. As reported in the lower panel of Figure~\ref{fig:C2H4_ImaginaryTime}, the convergence is fast also for higher values of $m$. Moreover, the asymptotic value of the energy that corresponds to the zero-point vibrational energy of ethylene agrees with the data reported in our previous work.\cite{Baiardi2017_VDMRG,Baiardi2019_HighEnergy-vDMRG}

\begin{figure}[htbp!]
	\centering
	\includegraphics[width=.75\textwidth]{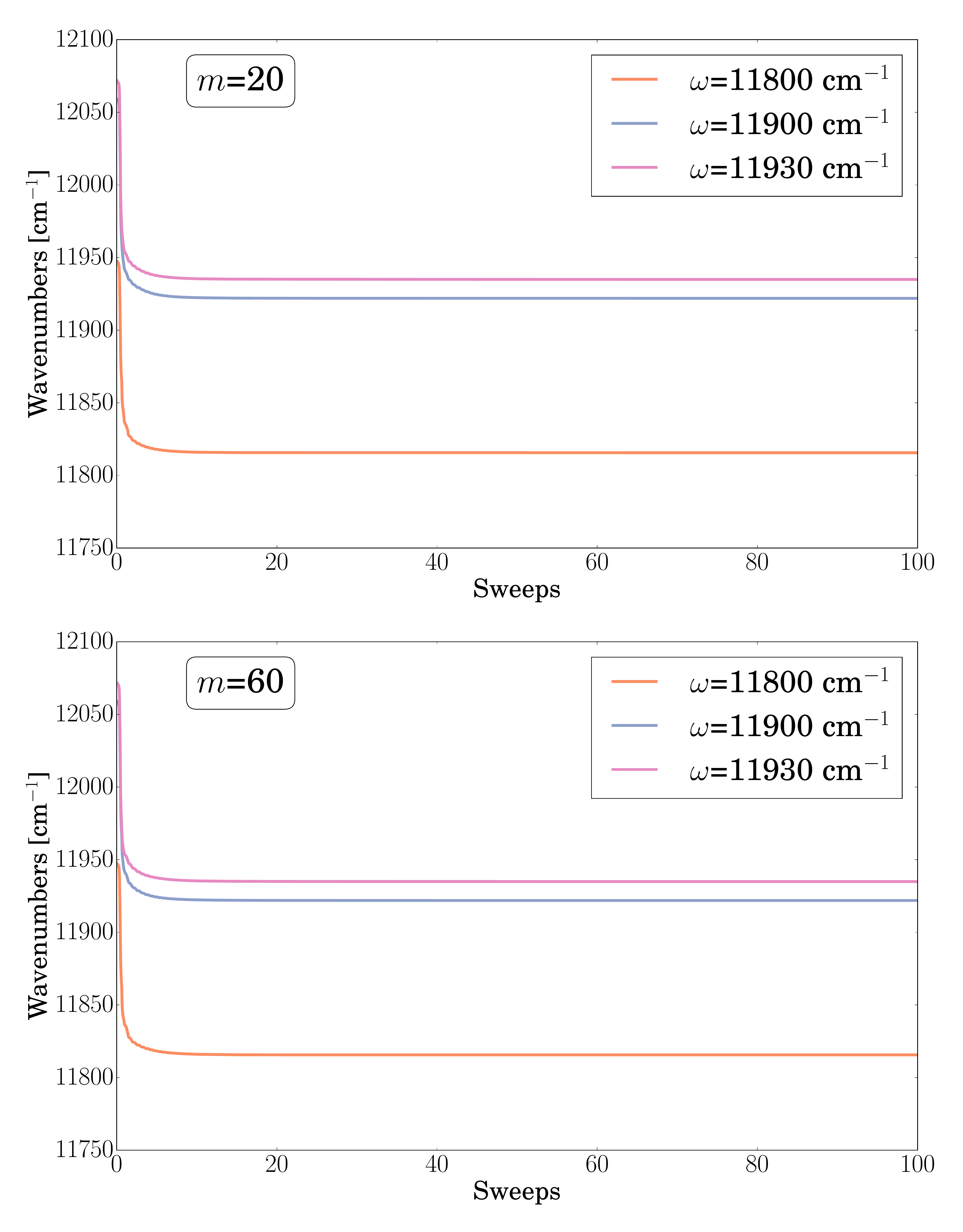}
	\caption{Energy of the three lowest vibrationally excited states of ethylene as a function of the number of sweep obtained with imaginary-time TD-DMRG. Calculations were performed with a time-step of 1~fs, N$_\text{max}$=6, $m$=20 (upper panel) and $m$=60 (lower panel) and different values for the energy shift $\omega$.}
	\label{fig:C2H4_ExcStates_mConv}
\end{figure}

We extend our previous analysis to low-lying vibrational excited states of ethylene through the imaginary-time evolution with the square-and-shift operator defined in Eq.~(\ref{eq:ImaginaryTimePropagation_ExcitedStates}). We set the energy shift to 11800~cm$^{-1}$ for the first state, to 11900~cm$^{-1}$ for the second one, and to 11930~cm$^{-1}$ for the third one. As shown in Figure~\ref{fig:C2H4_ExcStates_mConv}, in all cases the optimization converges smoothly within several sweeps to the energy of the excited state, which is in good agreement with our reference data reported in Ref.~\citenum{Baiardi2019_HighEnergy-vDMRG}.

We have already noted in the context of our recently introduced time-independent energy-specific formulation of vDMRG\cite{Baiardi2019_HighEnergy-vDMRG} that an estimate of the excitation energy is required to choose the energy shift to ensure that the optimization converges to the desired excited state. The choice of $\omega$ is particularly critical in regions with a high density of states, in which root-flipping events can occur. In time-independent simulations, we employed a root-homing criterion (also known as maximum overlap\cite{Kammer1976_RootHoming,Reiher2003_ModeTrackingNanotubes,Reiher2004_ModeTracking,Gill2009_SCF-ExcitedStates}) to consistently follow, within an iteration of the optimization, the solution with the largest overlap with a predetermined MPS. Root-homing techniques cannot be adopted in imaginary-time propagation because only one state is available in each sweep. Nevertheless, it is possible to exploit a partially converged time-independent DMRG calculation to build a reliable guess for the energy and the wavefunction of the target state to be then employed in an imaginary-time propagation scheme.

\section{Conclusions}
\label{sec:conclusion}

In the present work, we extended the tangent space-based formulation of TD-DMRG\cite{Haegeman2016_MPO-TDDMRG} and applied it to the simulation of the vibrational and vibronic dynamics of molecular systems. Due to the intrinsic complexity of the Hamiltonians describing vibrational and vibronic motion in molecules, we chose an MPS/MPO approach\cite{Haegeman2016_MPO-TDDMRG} considered so far only for studying the dynamics of one-dimensional Hamiltonians.\cite{Ren2018_TDDMRG-Temperature,Kurashige2018_MPS-MCTDH} Our approach has several advantages over alternative TD-DMRG schemes. First, it supports Hamiltonians with general structures containing both short- and long-range interactions. Moreover, it leads to a set of coupled differential equations that can be integrated exactly. This is achieved by constraining the time evolution of the wavefunction on the manifold composed by MPSs with a given rank. We extend the range of applicability of our TD-DMRG formulation by two alternative strategies to dynamically adapt the dimension of the MPS during the propagation. The first one employs a subspace-expansion scheme\cite{McCulloch2015_Mixing} and the other one relies on a two-site integration.\cite{Legeza2003_DynamicalBlockState} We first investigated the reliability of our implementation on the excitonic Hamiltonian describing the excitation energy transfer in a perylene bisimide aggregate. The application of TD-DMRG for a 24-mode vibronic Hamiltonian, describing the S$_1$ and S$_2$ electronic states of pyrazine, demonstrated the reliability of the algorithm when applied to large systems, which are difficult to access with current quantum dynamics approaches. We further generalized our TD-DMRG formulation to imaginary-time propagation, which represents an alternative to the standard, sweep-based scheme for optimizing ground and excited states as MPSs. 

On the one hand, the results of the simulations suggest that the bond dimension of an MPS should be increased constantly during the propagation to keep the truncation error, \textit{i.e.} accuracy of the MPS representation, fixed for the whole propagation. On the other hand, our tests show that employing a fixed value for the bond dimension $m$, and therefore letting the truncation error grow with time, enables one to obtain accurate results. This is probably due to the fact that absorption spectra are mainly determined by short-time dynamics and, therefore a small value of the bond dimension is sufficient to represent them accurately. However, additional investigations are required to check if this hypothesis is correct.

Due to its flexibility, our framework can be easily extended to the electronic Hamiltonian, the only additional hurdle being the inclusion of the spin symmetry. This hurdle can be overcome based on the strategies developed for standard time-independent DMRG.\cite{Keller2015_MPSMPODMRG,Chan2017_SpinAdapted-DMRG} The resulting theory represents a cost-efficient alternative to time-dependent CI\cite{Mazziotti2010_TDCI,Sonk2011_TDCI,Li2018_RealTime-GUGA_CI,Peng2018_TDCI} and allows one to simulate the electron dynamics for systems that are currently targeted by approximate approaches such as real-time time-dependent density functional theory.\cite{VanVoorhis2006_RealTimeTDDFT,Lopata2011_RealTimeTDDFT,Li2018_RealTime-Review} In the present formulation of the TD-DMRG algorithm, the local basis set is fixed during the whole propagation. A simultaneous optimization of both the CI coefficients, expressed as MPS, and of the local basis in a self-consistent-field fashion could enhance the efficiency of TD-DMRG. If applied to the electronic Hamiltonian, the resulting theory will constitute a cost-effective alternative to the time-dependent complete active space selft-consistent-field algorithm recently proposed in the literature,\cite{Sato2013_TD-CAS,Madsen2014_TDCASSCF_Singles,Sato2016_Atoms-RealTimeCASSCF} applied only to atoms and few-electron systems so far. For vibrational Hamiltonians, the coupled optimization of CI coefficients and local basis functions will lead to a DMRG approach analog of MCTDH.\cite{Bonfanti2018_MCTDH-ProjectorSplitting,Kurashige2018_MPS-MCTDH}

Besides being a cost-effective alternative to MCTDH, TD-DMRG can also drive the definition of strongly-interacting degrees of freedom for ML-MCTDH.\cite{Meyer2013_MLMCTDHVibronic,Wang2015_ML-MCTDH} For electronic structure theory, entanglement-based measures\cite{Legeza2003_OrderingOptimization,Legeza2011_Entanglement-DifferentStructures,Boguslawski2012_OrbitalEntanglement,Tecmer2015_OrbitalEntanglement,Stein2016_AutomatedSelection,Stein2017_OrbitalEntanglment} obtained from DMRG calculations have been demonstrated to be reliable metrics to detect strong-correlation between orbitals. When generalized to vibrational wavefunctions,\cite{McKemmish2011_VibronicEntanglement,McKemmish2015_VibronicEntanglement,Izmaylov2017_Entanglement} the same metrics could support the identification of strongly interacting vibrational degrees of freedom.

All results reported in the present work are obtained from dynamics performed on fitted potential-energy surfaces obtained as a Taylor expansion around a reference geometry. This is a major limitation, especially for long-time dynamics in which the wavefunction explores regions of the potential energy surface that cannot be efficiently represented by a power series in terms of some coordinate set. Some attempts are available in the literature to perform on-the-fly dynamics with MCTDH.\cite{Haberson2017_MCTDH-MLPotential,Haberson2019_DirectDynamics-FittedPES} These algorithms find a natural extension in DMRG through the framework discussed in this work.

\begin{acknowledgement}
This work was supported by ETH Z\"{u}rich through the ETH Fellowship No. FEL-49 18-1.
\end{acknowledgement}

\providecommand{\latin}[1]{#1}
\makeatletter
\providecommand{\doi}
  {\begingroup\let\do\@makeother\dospecials
  \catcode`\{=1 \catcode`\}=2 \doi@aux}
\providecommand{\doi@aux}[1]{\endgroup\texttt{#1}}
\makeatother
\providecommand*\mcitethebibliography{\thebibliography}
\csname @ifundefined\endcsname{endmcitethebibliography}
  {\let\endmcitethebibliography\endthebibliography}{}

\end{document}